\documentclass[%
aip,
amsmath,amssymb,
reprint,%
]{revtex4-1}

\usepackage{array}
\usepackage{graphicx}	
\usepackage{epsfig}
\usepackage{amsmath}		
\usepackage[colorlinks=false,linkcolor=blue]{hyperref}
\usepackage{xcolor, soul} 

\begin{document}

\title{Extreme Events in the Higgs Oscillator: A Dynamical Study and Forecasting Approach}


    \author{Wasif Ahamed M} 
    \email{ahamedw019@gmail.com}
	\affiliation{PG \& Research Department of Physics, Nehru Memorial College (Autonomous), Affiliated to Bharathidasan University, Puthanampatti, Tiruchirappalli - 621 007, India.}
    \affiliation{Department of Nonlinear Dynamics, School of Physics, Bharathidasan University, Tiruchirappalli - 620 024, India.}

    \author{Kavitha R} 
    \email{kavishiva2013@gmail.com}
	\affiliation{PG \& Research Department of Physics, Nehru Memorial College (Autonomous), Affiliated to Bharathidasan University, Puthanampatti, Tiruchirappalli - 621 007, India.}

    \author{Chithiika Ruby V}
    \email{chithiikaruby.v@trp.srmtrichy.edu.in}
    \affiliation{Center for Nonlinear and Complex Networks, SRM TRP Engineering College, Tiruchirappalli-621 105, Tamil Nadu, India.}
	\affiliation{Center for Research, Easwari  Engineering College, Chennai-600 089, Tamil Nadu, India.}
         
	\author{Sathish Aravindh M} 
    \email{sathisharavindhm@gmail.com}
	\affiliation{Department of Nonlinear Dynamics, School of Physics, Bharathidasan University, Tiruchirappalli - 620 024, India.}
	
		\author{Venkatesan A}
    \email{av.phys@gmail.com}
	\affiliation{PG \& Research Department of Physics, Nehru Memorial College (Autonomous), Affiliated to Bharathidasan University, Puthanampatti, Tiruchirappalli - 621 007, India.}
		
	\author{Lakshmanan M}
	\email{lakshman.cnld@gmail.com}
	\affiliation{Department of Nonlinear Dynamics, School of Physics, Bharathidasan University, Tiruchirappalli - 620 024, India.}

\begin{abstract}
Many dynamical systems exhibit unexpected large amplitude excursions in the chronological progression of a state variable. In the present work, we consider the dynamics associated with the one-dimensional Higgs oscillator which is realized through gnomic projection of a harmonic oscillator defined on a spherical space of constant curvature onto a Euclidean plane which is tangent to the spherical space. While studying the dynamics of such a Higgs oscillator subjected to damping and an external forcing, various bifurcation phenomena, such as symmetry breaking, period doubling, and intermittency crises are encountered. As the driven parameter increases, the route to chaos takes place via intermittency crisis and we also identify the occurrence of extreme events due to the interior crisis. The study of probability distribution also confirms the occurrence of extreme events. Finally, we train the Long Short-Term Memory neural network model with the time-series data to forecast the extreme events (EEs). 
\end{abstract}

\maketitle

\begin{quotation}
This study provides new insights into the dynamics of non-polynomial oscillators by investigating a one-dimensional Higgs oscillator under the effects of damping and external driving. We examine the conditions that give rise to extreme events - sudden, large - amplitude oscillations - within this nonlinear system. Without the influence of damping and forcing,  it exhibits simple harmonic oscillations with  amplitude dependent frequency. The presence of damping and external driving allows us to explore how nonlinear phenomena emerge in the system, revealing behaviors that differ from simple oscillatory motion. This research deepens our understanding of the unique and complex behaviors of non-polynomial systems. Through comprehensive analyses, including time series, bifurcation diagrams, and Poincar\'e sections, we identify the pathways to these rare extreme events. Statistical methods further confirm the rarity of such occurrences, while a Long Short-Term Memory (LSTM) neural network model is applied to forecast extreme events, demonstrating potential predictive applications. 
\end{quotation}

\section{Introduction}
Nonlinear oscillators, as defined on a Euclidean plane and introduced by Mathews and Lakshmanan \cite{mathews1974unique, mathews1975quantum} and Higgs \cite{higgs1979dynamical}, can be associated with the harmonic oscillator on a spherical space. This association is established through orthogonal and gnomic projections onto the Euclidean plane, which is considered to be tangent to the spherical space. They exhibit simple harmonic oscillatory behavior. Mathews and Lakshmanan nonlinear oscillator and its various generalizations are continuously studied for many aspects such as $3$-dimensional \cite{lakshmanan1975quantum} and $d$-dimensional generalizations as well as quantum aspects  \cite{ranada2002harmonic, lakshmanan2013generating} and rational extensions of the potentials  \cite{quesne2018deformed}. Similarly, the Higgs oscillator has received significant attention since its introduction in the literature, particularly with a focus on the concept of superintegrability \cite{ballesteros2008superintegrable, ruby2021classical,V_2024}. Additionally, remarkable studies have focused on the examination of the cuboctahedric Higgs oscillator \cite{hakobyan2009cuboctahedric}, investigations into the quantum dynamics of its relativistic generalization \cite{mohammadi2016dirac}, exploration of its quantum exact solvability \cite{carinena2004non}, the formulation of nonlinear coherent states involving the observation of deformed oscillator algebra \cite{mahdifar2006geometric}, and the exploration of hidden symmetries and the conformal algebra \cite{evnin2016ads}.

Understanding the dynamics of damped-driven nonlinear systems helps researchers to predict responses to stimuli, optimize performance, and design adaptable systems for varying environmental conditions. Different nonlinear phenomena were shown to be exhibited by Mathews and Lakshmanan oscillator when it is subjected to damping and driven forces \cite{venkatesan1997nonlinear}. Motivated by this work, we are interested to understand the dynamics of the Higgs oscillator under the combined effect of damping and driven forces.  Within this context, the system is shown to exhibit various bifurcation phenomena, including symmetry breaking, period doubling, and intermittency bifurcations. Further, it shows extreme behavior through the interior crisis-induced intermittency as the strength of the driven parameter increases. It has been reported in the literature that intermittency route is a major pathway for the emergence of extreme events in chaotic systems \cite{ott2002chaos}. Crisis-induced intermittency \cite{grebogi1983crises, grebogi1987critical} and Pomeau-Manneville (PM) intermittency \cite{pomeau2017intermittent} have been reported to be associated with the occurrence of extreme events \cite{zamora2013rogue, kingston2017extreme, mishra2020routes, chowdhury2022extreme,kaviya2023route}. Such extreme behavior is observed in dynamical systems such as the Li\'enard system \cite{kingston2017extreme}, FitzHugh-Nagumo oscillators \cite{saha2017extreme}, the Hindmarsh-Rose model \cite{saha2018riddled, vijay2023superextreme},  electronic circuits \cite{kingston2017extreme, thangavel2021extreme}, network of Josephson junction \cite{ray2020extreme}, the nonlinear Schr\"odinger equation \cite{fotopoulos2020extreme}, and dynamical systems with discontinuous boundaries \cite{kumarasamy2018extreme}. {Also, the emergence of large-amplitude events has been observed in the rotational dynamics of a damped pendulum under the influence of both dc and ac torques \cite{pal2023extreme}.}  The observed extreme events are mitigated by applying  an additional constant source (bias) \cite{sudharsan2021constant}. The mitigation technique helps to control the emergence of extreme events in dynamical systems \cite{bonatto2017extreme, sudharsan2021emergence, sudharsan2022suppression, chowdhury2022extreme}.

Previously several authors have studied the extreme events which occur in polynomial oscillators \cite{kingston2017extreme,kumarasamy2018extreme,kaviya2023route,roy2022model,saha2017extreme,saha2018riddled,chowdhury2022extreme,shashangan2024mitigation}. But the exploration of extreme events in nonpolynomial  oscillators is significantly necessary for the reason that many physical systems exhibit oscillatory behavior that cannot be described by polynomial functions. Examples are pendulum, mechanical systems, electronic circuits, and some biological rhythms. Understanding these non-polynomial oscillators is crucial for optimizing performance and predicting dynamical behavior in various physical systems. 
{Mainly our research focuses on a nonlinear oscillator, known as the Higgs oscillator, where the nonlinearity comes from the curvature of the underlying space, described by a position-dependent, non-Euclidean metric. Most of the studies on chaotic behavior in nonlinear oscillators focus on systems within a flat space, incorporating damping and driving forces, whereas the present work emphasizes the role of spatial curvature in shaping the dynamics. This leads to the emergence of chaotic behavior under driven and damped conditions. The detailed analysis of the one-dimensional system reveals various bifurcations, critical transitions, extreme events, and chaos which set a foundation for exploring more complex dynamics in higher-dimensional systems in non-Euclidean spaces. }

{Position-dependent mass (PDM) systems are extensively studied in quantum mechanics \cite{bastard1990wave, gonul2002exact}. Our previous investigations established a novel connection between PDM systems and quadratic Li\'enard-type nonlinear  systems, providing a fresh perspective on their classical dynamics \cite{mathews1975quantum, ruby2021classical, V_2024}. In this manuscript, we extend this framework by exploring the classical dynamics of PDM systems under the influence of damping and driving forces. This approach not only enriches the understanding of classical nonlinear dynamics in PDM systems but also lays a foundational framework for future studies on how damping and driving conditions in classical systems might translate into the quantum regime.} Also, to the best of our knowledge, it is for the first time the non-integrable  dynamics of  the damped-driven one-dimensional Higgs oscillator has been analyzed in the literature.

In recent years,  several machine learning (ML) algorithms have been utilized for  performing  different  cognitive tasks like pattern recognition, classification, prediction, natural language processing,etc., for automation in industrial processes and in several scientific research works \cite{jordan2015machine, mohri2018foundations,  carleo2019machine, alpaydin2020introduction, alpaydin2021machine}. It is quite notable that ML Algorithm also gains much potential in solving nonlinear dynamical equations \cite{ljung2010approaches, brunton2016discovering, tang2020introduction, pourmohammad2023dynamical}. 

Mathematical modeling of dynamical systems and understanding the dynamical behavior from limited empirical data is a tedious task. To overcome such difficulties, ML algorithms like recurrent neural network (RNN) \cite{pan2011model}, gated recurrent unit (GRU) \cite{liu2023nonlinear}, long short-term memory (LSTM) network \cite{meiyazhagan2021model, chen2022intelligent, meiyazhagan2021prediction, chowdhury2022extreme} and higher RNN based echo state network have been widely employed for predicting the dynamics of the system without explicit models. Prior studies have shown that LSTM  achieves  better prediction accuracy for long Lyapunov times in temporal tasks \cite {meiyazhagan2021model} as it has the memory of the past history of states for prediction. Using the forementioned  machine learning algorithms (ML), predicting time series, temporal patterns in coupled dynamical systems and spatio-temporal patterns  \cite{jaeger2004harnessing, pathak2021reservoir, BARMPARIS2020126300, Kushwaha_2021,ganaie2020identification} are well established. Also, it is quite notable that parameter-aware ML algorithms were developed to identify the bifurcation parameter of the system and the associated dynamics \cite{xiao2021predicting} along which it also predicts the basins of attraction in multistable system \cite{roy2022model}. {The mathematical optimization-based ensemble deep learning model, referred to as the Optimized Ensemble Deep Learning (OEDL) framework, is designed to forecast extreme events using both mathematical models and real-time data, such as COVID-19 statistics and climate information \cite{ray2021optimized}.}

One of the peculiar dynamical behavior is the occurrence of rare high amplitude extreme event (EE) oscillations. Prediction and mitigation of such rare events prior to its occurrence have gained increasing interest recently \cite{shashangan2024mitigation}. Using ML algorithms several authors  have predicted the extreme rare events in time series prior to its  emergence. Using these  ML algorithms, EEs have been forecasted in coupled neuronal system \cite{PYRAGAS2020126591}, mechanical systems \cite{meiyazhagan2021model, durairaj2023prediction}, complex Ginzburg-Landau system \cite{jiang2022predicting}, etc. Here, we have used the LSTM model to forecast the EEs associated with the chaotic dynamics of the Higgs oscillator.  

In this paper, we plan to present a detailed study of the damped, driven, Higgs oscillator where we have investigated the existence of a rich variety of bifurcations as a function of the external forcing parameter while all other system parameters are fixed. When the strength of the external force increases, transition to chaos takes place and extreme events emerge in the dynamical system. Extreme behavior has been characterized by specific bifurcation and probability distribution function. Finally, the LSTM model is trained with chaotic data related to extreme event. Then we test the model by predicting the future time series capturing the transitions to EEs from bounded chaotic motions. In addition, we used the mean square error and correlation plot to validate the prediction accuracy of the LSTM model.  

The structure of this paper is as follows: In section II, we discuss the classical solvability of the nonlinear system, and then analyze the dynamics involved in the Higgs oscillator with the inclusion of damping and external forcing terms in section III. Subsequently in sections IV and V, we discuss the mechanism for EE and characterization of the EE, respectively. In section VI, we demonstrate the forecasting of the time series of EEs using the LSTM machine learning algorithm. Finally, we conclude by summarizing our findings in section VII.

\section{ONE DIMENSIONAL HIGGS OSCILLATOR}
The classical nonlinear oscillator introduced by Mathews and Lakshmanan is characterized by the non-polynomial Lagrangian,
\begin{equation}
    L = \dfrac{\dot{x}^2-\omega_{0}^{2}x^{2}}{2(1+\lambda x^{2})},
    \label{eq11}
\end{equation}
where $\omega_{0}$ and $\lambda$ are parameters. It has drawn considerable interest in the literature due to its unique simple harmonic oscillatory behavior \cite{mathews1974unique,mathews1975quantum}. Venkatesan and Lakshmanan have studied the dynamics of the nonlinear system subjected to linear damping and periodic forcing \cite{venkatesan1997nonlinear}. In particular, the authors have shown that the system exhibits a rich variety of bifurcation phenomena. It includes the familiar period-doubling bifurcation phenomena, preceded by a symmetry-breaking bifurcation. In addition to the period-doubling route to chaos, intermittency, and antimonotonicity have been found to exist in the mechanical system. Also, they confirmed the existence of torus doubling followed by strange nonchaotic attractor \cite{venkatesan1997nonlinear}. 
 Recently, two of the present authors have studied the classical and quantum solvability of the one-dimensional version of the Higgs oscillator \cite{ruby2021classical}. In the present work, we are also interested in understanding the dynamics of the one-dimensional Higgs oscillator under the combined effects of damping and driven forces. The non-linear system can be geometrically interpreted as it arises from the generalization of Cartesian coordinates in Euclidean geometry \cite{ruby2021classical}. In the one-dimensional case, the Lagrangian takes the form:
\begin{equation}
	L = \dfrac{\dot{x}^2}{2(1+\kappa x^2)^2}-\dfrac{\omega_0^2 x^2}{2},
	\label{eq2}
\end{equation}
where $\kappa$ is a parameter related to the curvature. The associated equation of motion \cite{ruby2021classical, V_2024},  
\begin{equation}	
\ddot{x}-\dfrac{2\kappa x}{(1+\kappa x^2)}\dot{x}^2+\omega_0^2(1+\kappa x^2)^2 x=0,~~~~~~~~~\big(^{.}=\dfrac{d}{dt}\big)
\label{equ1}
\end{equation}
 can be exactly integrated, and  its solution has been expressed in terms of trigonometric functions as
\begin{equation}
x(t)=\dfrac{A \sin(\Omega t + C)}{\sqrt{1-\kappa A^2 \sin^2(\Omega t + C)}},
\label{equ5}
\end{equation}
where $ A $ and ${\displaystyle \Omega=\frac{\omega_{o}}{\sqrt{1-\kappa A^{2}}}}$ are the amplitude and frequency of the oscillator, respectively, while C is an arbitrary phase constant. Hence, $x(t) $ is periodic for all the values of $ A $ for $ \kappa < 0 $ and periodic in the range $ |A|<\dfrac{1}{\sqrt{\kappa}} $ when $ \kappa >0 $. 

To study the nonlinear dynamics of the damped and driven version of the system, we consider the one-dimensional Higgs oscillator, under the influence of additional damping and external periodic forcing so that the equation of motion becomes
\begin{equation}
\ddot{x}-\dfrac{2\kappa x}{(1+\kappa x^2)}\dot{x}^2+\omega_0^2(1+\kappa x^2)^2 x+\alpha \dot{x}  = f \cos \omega t.
\label{equ6}
\end{equation}
Here $\alpha$ is the damping parameter, $ f $  and $\omega$ are the amplitude and frequency of the external driving force. For the purpose of numerical analysis we have rewritten eq.\eqref{equ6} into two first order equations,  
\begin{subequations}
      \begin{eqnarray}
     \dot{x} &=& y  \label{eq:6a},\\
     \dot{y} &=& \dfrac{2\kappa x}{(1+\kappa x^2)}y^2 -\omega_0^2(1+\kappa x^2)^2 x-\alpha y  + f \cos \omega t. ~~~~~  \label{eq:6b}
    \end{eqnarray}
\end{subequations}

For numerical integration, we have used the Runge-Kutta fourth order with fixed step size of $ h = \frac{2\pi}{2000\omega} $. In particular, we will investigate the dynamics of the system and the existence of a rich variety of bifurcations, as the external forcing parameter is varied while keeping all other parameters fixed. 

\section{Higgs Oscillator with damped and driven force}
\subsection{Bifurcation studies}
\begin{figure}
\centering
\includegraphics[width=0.8\linewidth]{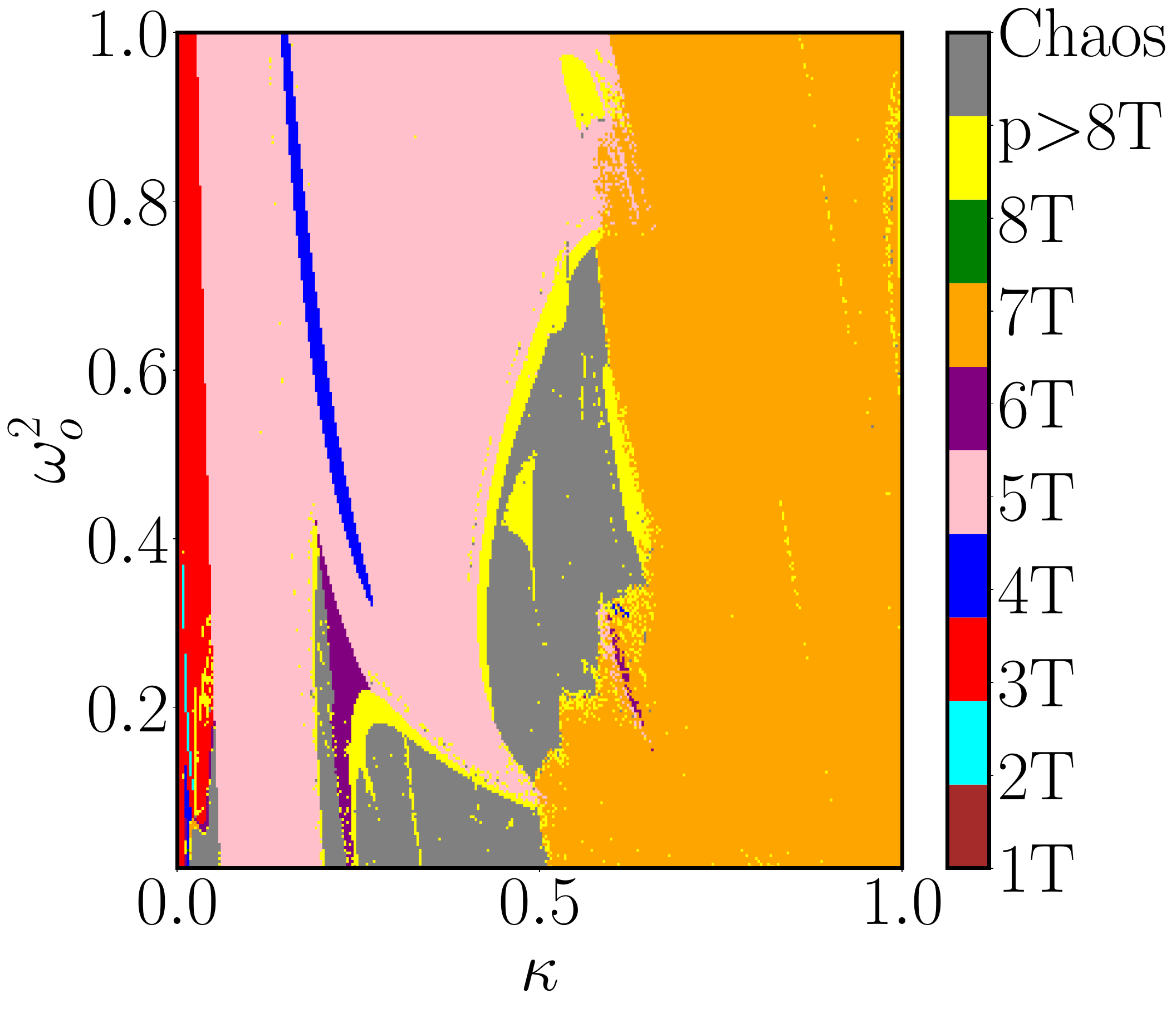}
\caption{The two-parameter ($\kappa$ vs $\omega^{2}_{o}$) phase diagram associate with system (\ref{equ6}) for identifying different transitions with fixed parameters $\alpha=0.05$ , $f=4.88463$, and  $\omega = 0.5$.}
\label{fig1_phase_space}
\end{figure}

The primary focus of this paper is to examine bifurcations and extreme events associated with system \eqref{equ6}. Before examining the occurrence of extreme events, we explore the dynamics of the Higgs oscillator subjected to the damping and driven forces. This is achieved by varying the system parameters and analyzing the results through bifurcation diagrams, time series plots, Poincar\'e sections, and Lyapunov exponents. Through investigation on the routes to chaos, it is identified that the system predominantly follows the period-doubling route and intermittency transitions to chaos. It is important to note that the parameters used here are specifically set to observe extreme events, with further details provided in the subsequent section.
 Initially, the values of the system parameters are chosen as $\alpha=0.05$ , $f=4.88463$, $\omega = 0.5$ and are fixed. In this section, various dynamical states are classified, and the tracing of the state transitions is shown by scanning the system parameters $\kappa$ vs. $\omega^{2}_{o}$. The resulting two-parameter phase diagram for the $\kappa$-$\omega^{2}_{o}$ parameters is shown in Fig.\ref{fig1_phase_space}. This diagram illustrates the various dynamical states that exist in the Higgs oscillator for the above choice of $\alpha$,$f$ and $\omega$. The two parameter-phase diagram uses different colors to represent specific time periods: brown, cyan, red, blue, pink, purple, orange, and green respectively correspond to 1T-8T oscillations, yellow denotes $p>8T$, and gray denotes chaos, with $p$ denoting the period of oscillations. In this diagram, we observe the regions where the gray chaotic zones are surrounded by yellow regions with $p>8T$, further enclosed by regions with $p<8T$. This pattern clearly signifies the period-doubling route to chaos. Even some chaotic regions share some of their boundaries directly with the $p<$8T region, which shows a type III intermittency transition. Having understood such routes to chaos, in the section III B we will explore the routes and emergence of extreme events when varying the forcing amplitude $f$.

\subsection{Bifurcation studies while varying the control parameter \textit{f}}

\begin{figure}
\centering
\includegraphics[width=0.48\linewidth]{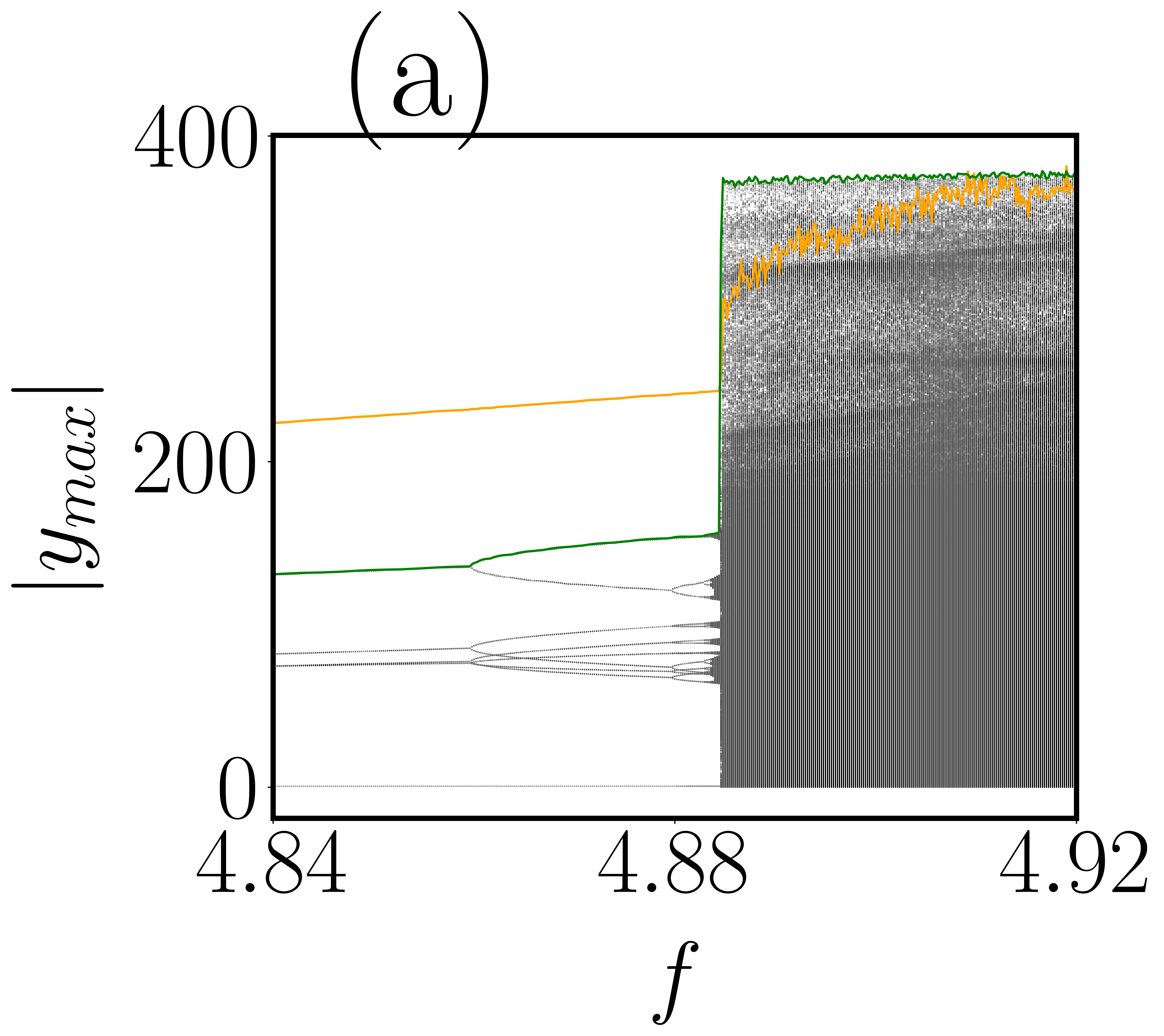}
\includegraphics[width=0.48\linewidth]{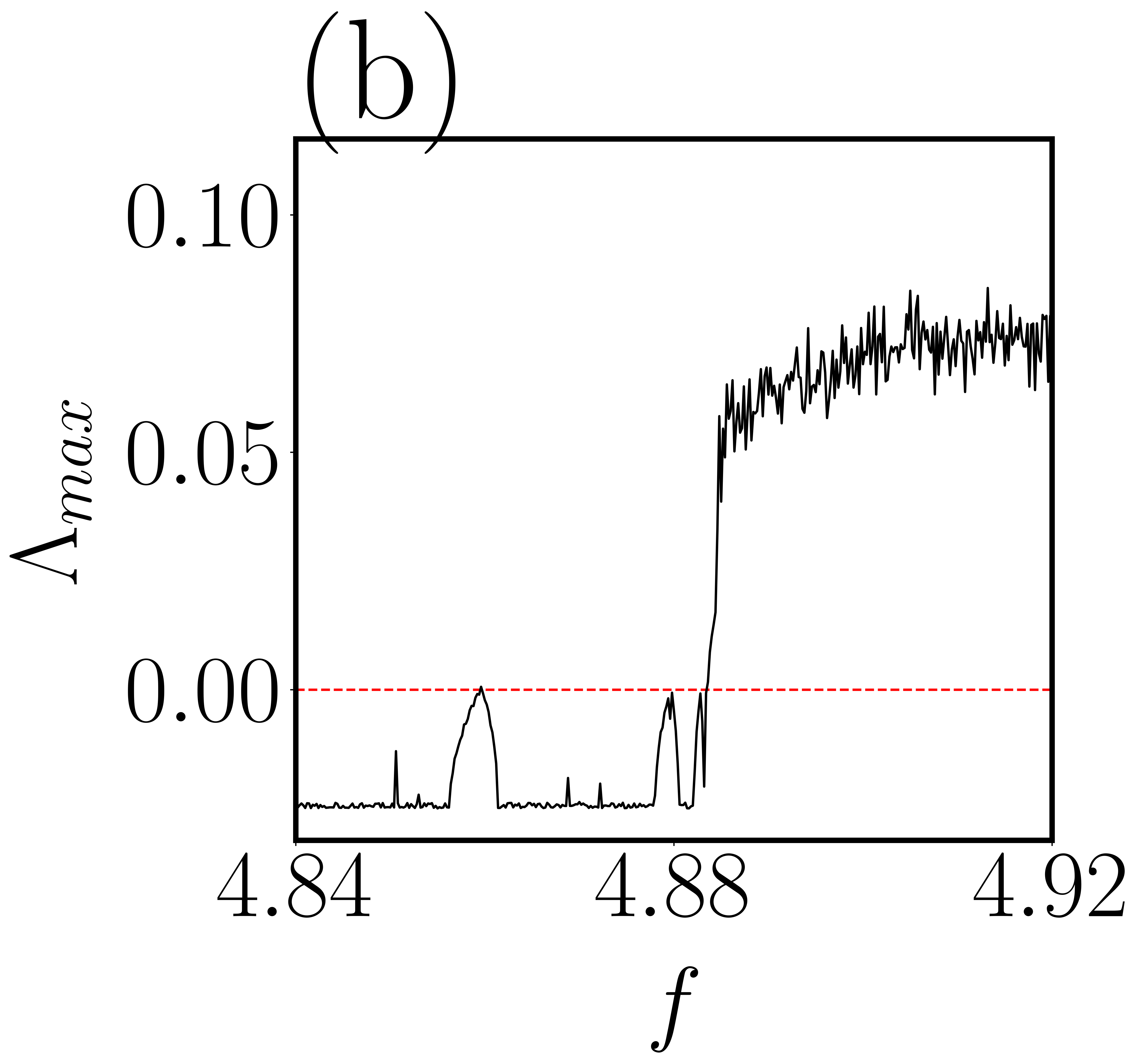}
\caption{{(a) shows the bifurcation diagram for $y(t)$ against the forcing parameter ($f$) in the range of 4.84 to 4.92. Here orange dotted line represent the extreme event threshold ($H_s$) of the dynamical state $y(t)$, respectively. (b) shows the largest Lyapunov spectrum for $\Lambda_{max}$ against the forcing parameter $f~ \in 4.84 : 4.92.$}}
\label{fig2}
\end{figure}

In this section, for the bifurcation analysis, we fix some of the parameters of the Higgs oscillator as mentioned in the previous section $\omega^{2}_{o} = 0.04$, $\kappa = 0.2$, $\alpha = 0.05$, and $\omega = 0.5$, and vary the forcing amplitude (\textit{f}). The bifurcation diagram shown in Fig.\ref{fig2}(a) is produced by plotting the maxima of asymptotic dynamical state $y(t)$ against the forcing parameter ($f$) in the range of 4.84 to 4.92. To clearly visualize the occurrence of extreme events (EEs) for a specific control parameter $f$, we label the green and orange points to represent the global maximum peak and the extreme event threshold ($H_s$) of the dynamical state $y(t)$, respectively.

The EE threshold ($H_s$) is numerically calculated from the time trajectories using the formula $H_s = \langle y_n \rangle + N\sigma(y_n)$, where $y_n$ is the peak of the time series signal of the state variable $y(t)$ considered for EE characterization. Here, $\langle y_n \rangle$ and $\sigma(y_n)$ represent the mean and standard deviation of the signal's maximum, and $N$ is the threshold qualifier parameter, which can take values such as 4, 5, 6, etc. In our analysis, we fix the parameter $N$ as 4.0.

In the bifurcation diagram Fig.\ref{fig2}(a), whenever the green points are above the orange points, it indicates that the trajectory crosses the threshold, signaling the onset of extreme events.
In the range of the values of $f\in(4.84 ,4.859)$,  period-5 orbit  is stable, and at $f=4.859$ the onset of the period  doubling  bifurcation occurs which is leading to  chaos. The unique property which distinguishes chaos from other periodic dynamics is its sensitive dependence on initial conditions. It is well known that the presence of infinite number of unstable periodic orbits (UPOs) embedded within the chaotic attractor is fundamentally responsible for its divergence properties \cite{grebogi1987unstable}.  This divergence property is quantitatively measured by calculating the maximum Lyapunov exponents ($\Lambda_{max}$) of the system \cite{wolf1985determining}. For chaos, $\Lambda_{max}$ is always positive. In Fig.\ref{fig2}(b), we could see the sudden transition from negative to positive $\Lambda_{max}$.
\begin{figure}
\includegraphics[width=1.0\linewidth]{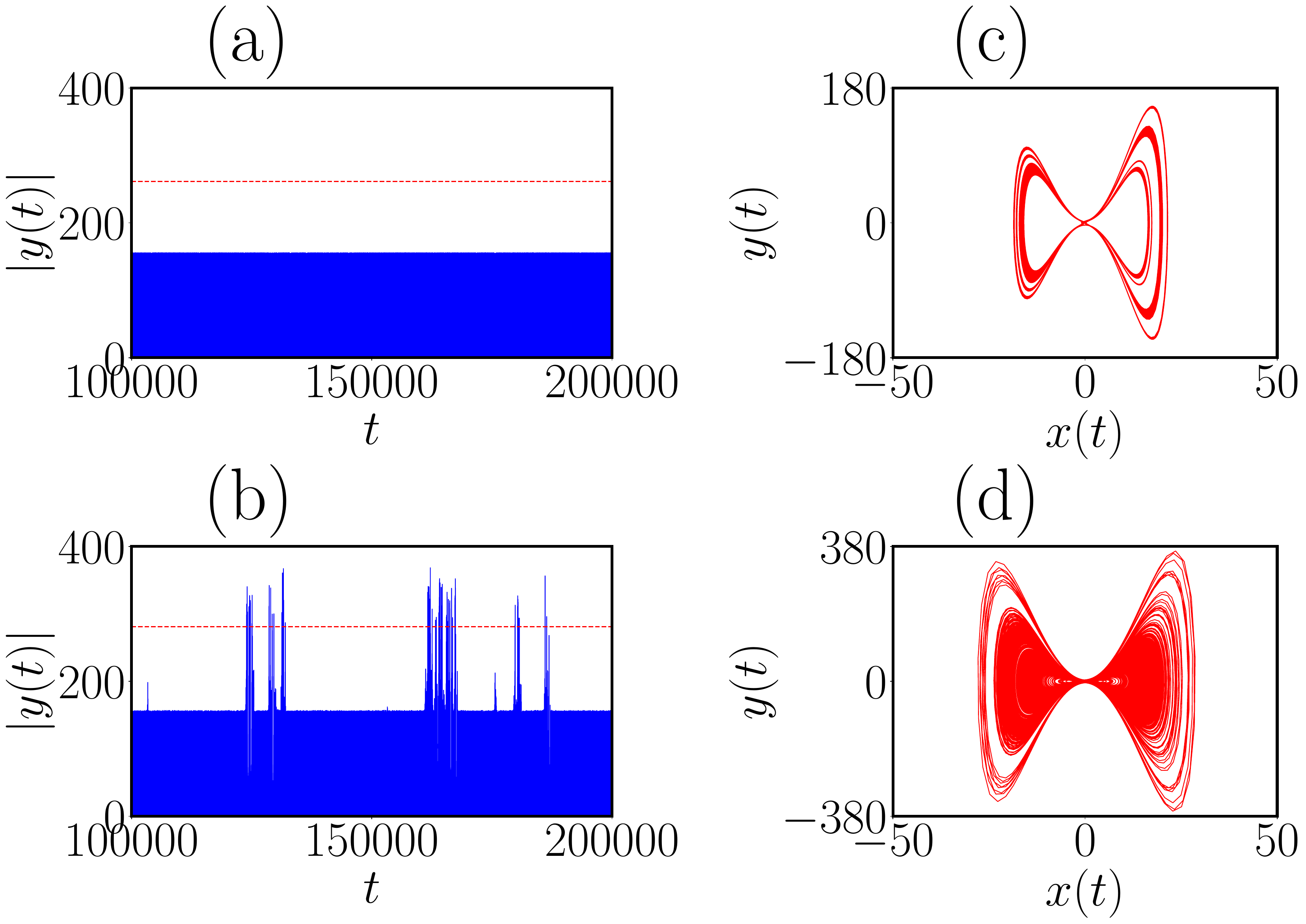}
\caption{ (a)-(b) $\&$ (c)-(d) show the absolute value of time series of the dynamical state variable $y(t)$ and the corresponding phase trajectories in the (x-y) plane,  for different external forcing parameter values, $ f =4.884$ and $ f=4.88463 $ in which other system parameters are fixed as $\omega^{2}_{o}=0.04$, $k=0.2$, $\alpha=0.05$ and $\omega=0.5$, respectively.} 
\label{fig3}
\end{figure}
In the region $4.88373< f<4.88463$, the chaotic trajectories are bounded under the EEs threshold ($H_{s}$) with $\Lambda_{max} > 0 $. At $\textit{f}=4.88463$, we could find the sudden expansion in the size of the attractor when the chaotic trajectory meets the chaotic saddles in the stable manifold  and there occurs a long excursion of trajectory (see Fig.\ref{fig2}). This  phenomenon is widely recognized as the interior crisis. Also, we observe the gradual increase in the size of the attractor on further increasing of \textit{f}.

\subsection{Time series and the observation of Extreme Events}

From previous studies in the literature\cite{chowdhury2022extreme}, it has been realized that EEs are caused due to few instabilities in phase space, namely the presence of saddle equilibrium points, singularities, and presence of saddle orbit.

\textbf{Boundary crisis}: The long excursion of trajectory occurs when chaotic trajectories collide with the basin boundary or with an unstable equilibrium point \cite{grebogi1982chaotic}.

\textbf{Interior crisis}: This type of intermittency  crisis  occurs when the chaotic trajectory meets the chaotic saddles and flows along the chaotic saddle orbit leading to long excursion in phase space at a critical parameter \cite{grebogi1982chaotic}. Generally the chaotic saddles are the regions which separate the stable and unstable manifolds. Moreover, the collision of the advancing period-doubling and period-adding cascades with the changes in the system parameter causes the interior crisis \cite{venkatesan1997nonlinear}. 

\textbf{Attractor merging crisis}:
This type of intermittency occurs when two or more chaotic attractors collide resulting in the formation of an expanded new chaotic attractor \cite{chossat1988symmetry}.

\textbf{Pomeau and Manneville Intermittency}:
In some dynamical systems, the periodic oscillations suddenly transits to the chaotic states via the saddle node bifurcation at some critical system parameter. This means that the dynamics is mostly periodic with interceded irregular intermittent chaotic bursts. It has been reported that these chaotic bursts in some scenarios trigger large amplitude EEs in some systems \cite{pomeau2017intermittent}.

From a detailed numerical analysis, we find that of the various bifurcation routes leading to EEs, our system \eqref{equ6} follows the interior crisis mechanism. To explore this in more detail, we analyze the time series and phase trajectory to demonstrate how extreme events emerge from the bounded chaotic attractor.
In the previous section, we used the bifurcation diagram in Fig.\ref{fig2}(a) to distinguish between regions of extreme event and non-extreme event, based on the extreme event threshold ($H_s$). After carefully examining the dynamics as a function of the control parameter $f \in (4.84,4.92)$ for the initial conditions $(x_{o},y_{o}) \in (0.11,0.22)$, and removing sufficient transients from the time series, we have identified (i) bounded chaos at $f = 4.884 $, where the time series remains below $H_s=261.405$ (via Fig. \ref{fig3}(a)) and (ii) extreme events at $f = 4.88463$, where the trajectory surpasses $H_s = 280.974$ (via Fig. \ref{fig3}(b)).

\begin{figure}
\centering
\includegraphics[width=0.8\linewidth]{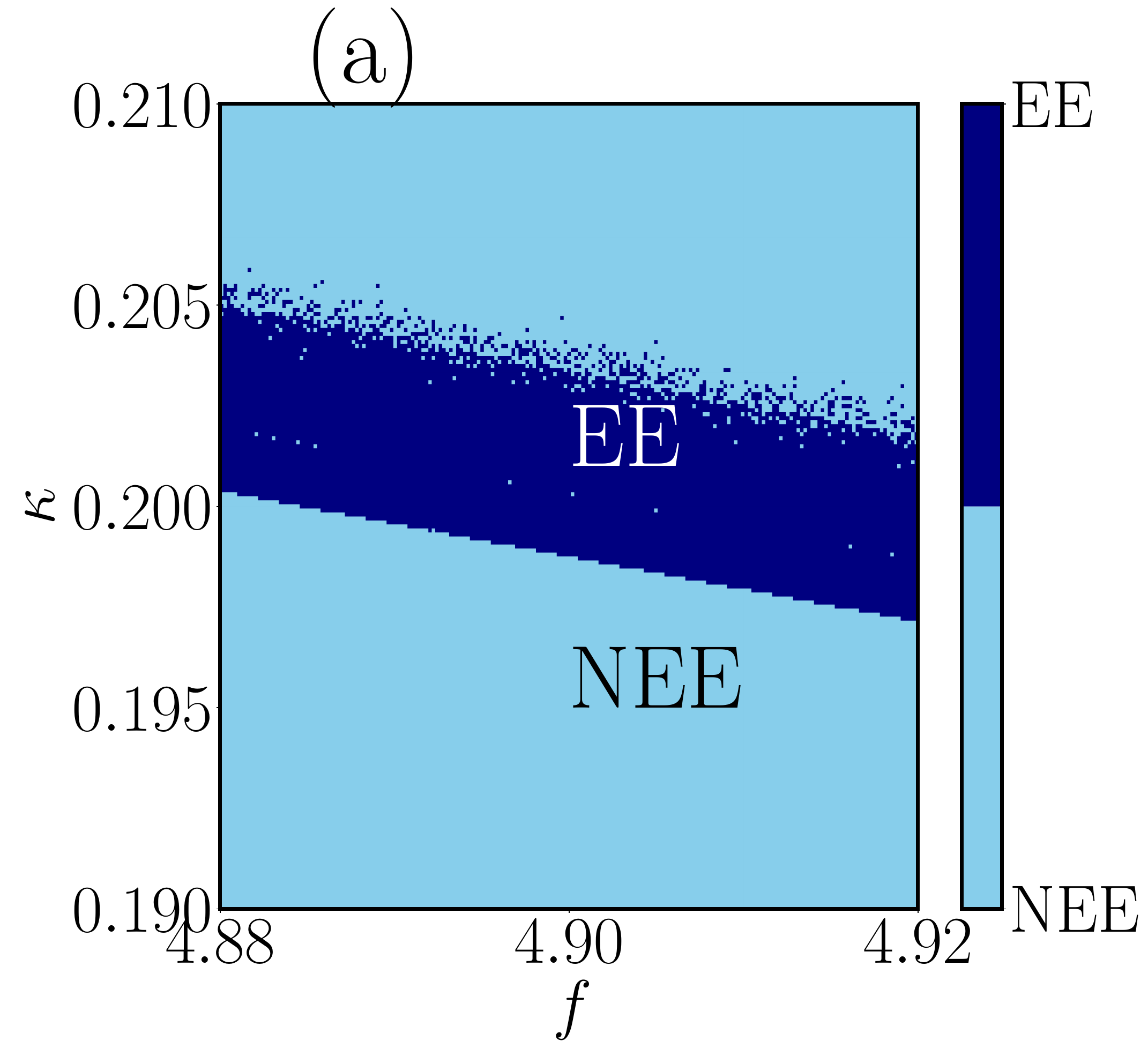}
\includegraphics[width=0.8\linewidth]{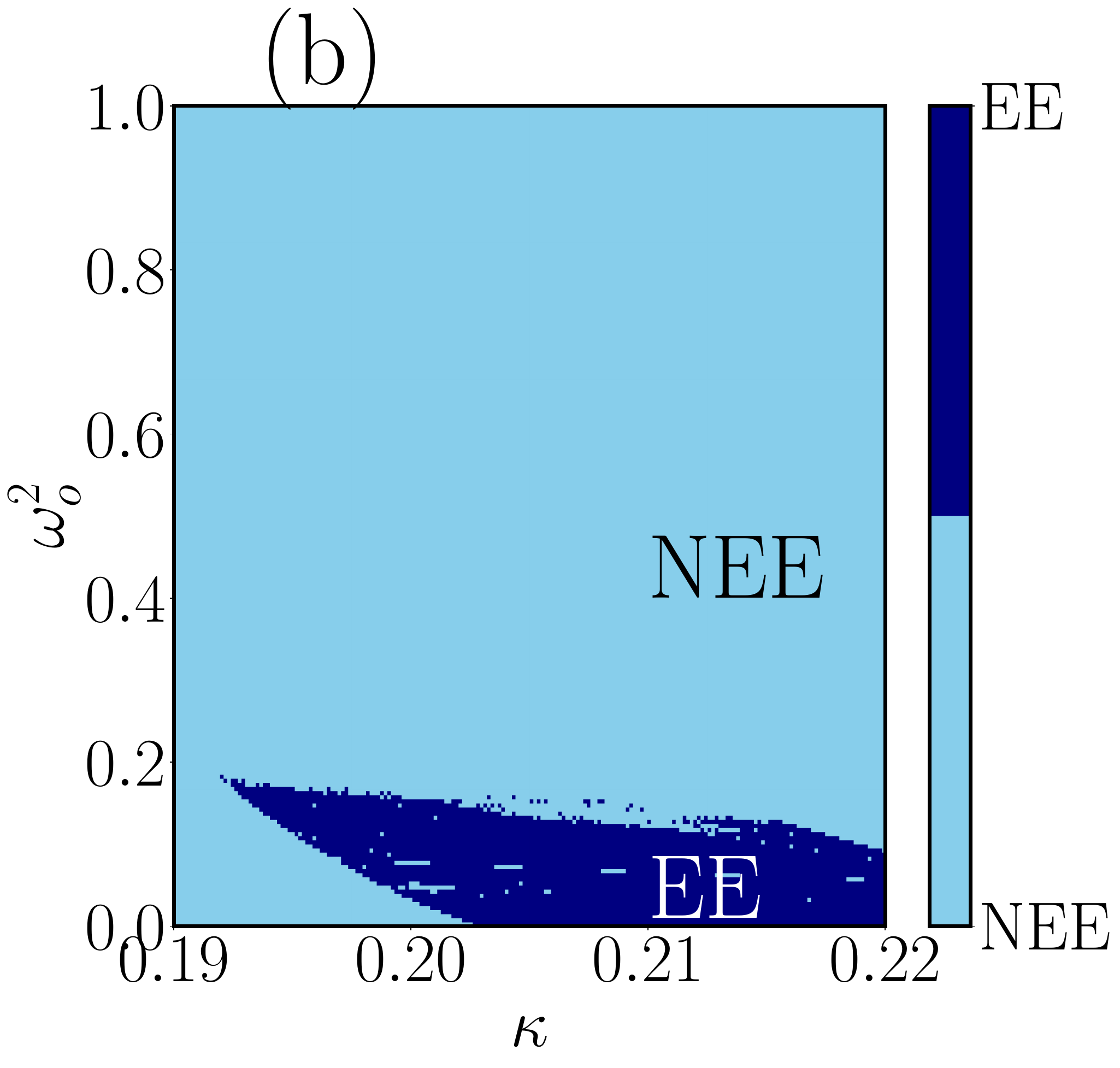}
\caption{(a)-(b) show the two-parameter phase diagram for ($f$ vs $\kappa$) and ($\kappa$ vs $\omega^{2}_{o}$) . The dark blue and light blue color regions represent the occurrence of EE and NEE, respectively.}
\label{fig4}
\end{figure}

We have generated a two-parameter phase diagram in Fig.\ref{fig4}(a) by fixing the parameter $f$ within the range $f \in (4.88, 4.92)$ and adjusting $\kappa$ in the range $\kappa \in (0.19, 0.21)$ to observe how the extreme events (EEs) associated with chaotic attractors change in the parameter space ($f$, $\kappa$). According to Fig.\ref{fig4}(a), EEs occur when $\kappa$ lies between 0.198 and 0.205, while the remaining regions exhibit non-extreme event (NEE) behavior.
Additionally, we have also produced a two-parameter phase diagram of ($\kappa$,$\omega^{2}_{o}$). As $\omega^{2}_{o}$ increases, the system becomes highly periodic and does not exhibit extreme events (NEE regions) when $\omega^{2}_{o} \in (0.2, 1.0)$ (see Fig.\ref{fig4}(b)). Below this range, periodicity decreases, and for lower values of $\omega^{2}_{o} \in (0.01, 0.2)$ and $\kappa \in (0.20, 0.22)$, the system exhibits EE behavior.
This study clearly demarcates the EE and NEE region in the parameter space and allows one to tune the system parameters suitably, which avoids the large amplitude extreme event behavior during system applications. 

\section{Stroboscopic maps and the mechanism of emergence of extreme events}
\begin{figure}
\centering
\includegraphics[width=0.8\linewidth]{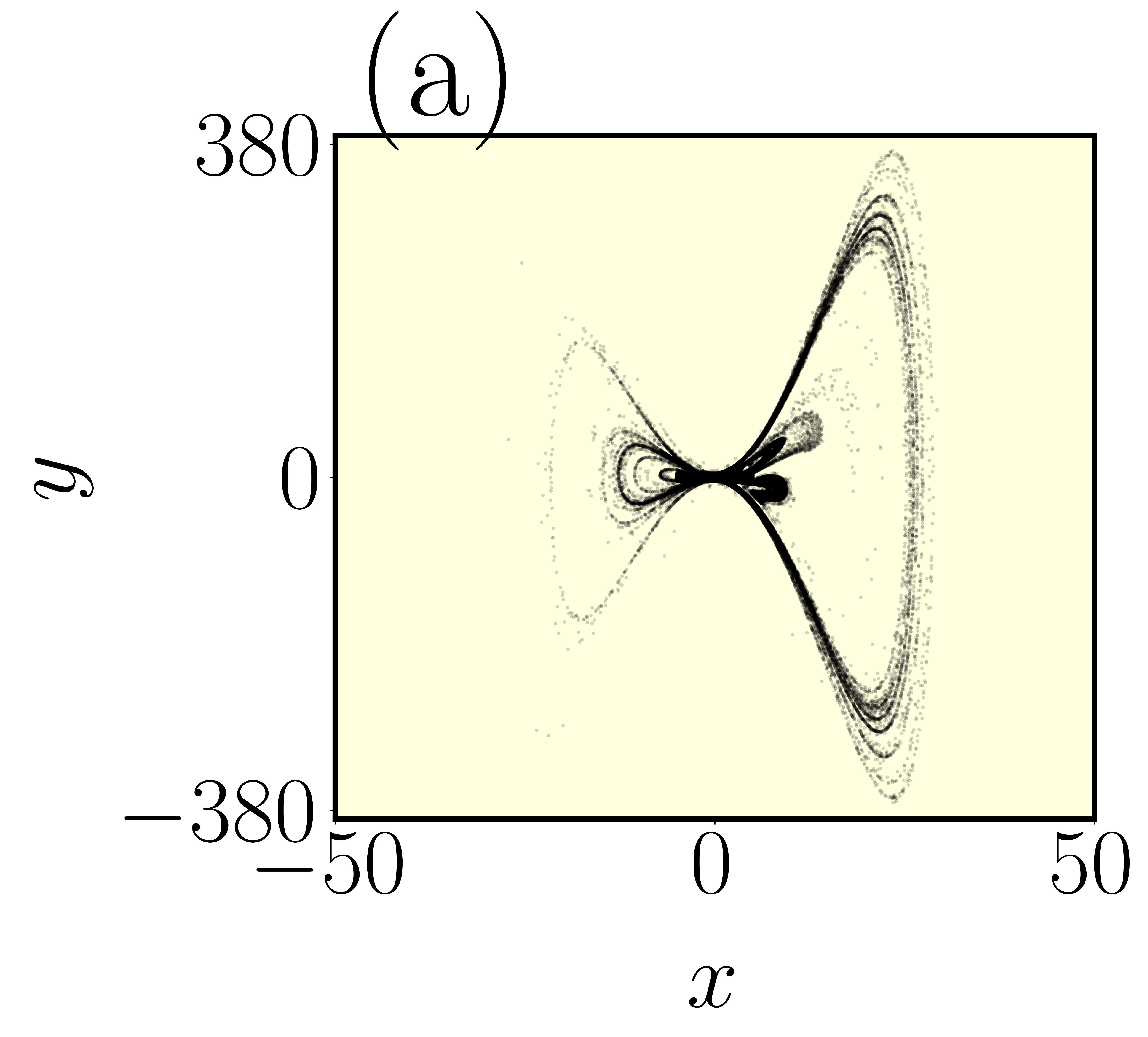}
\includegraphics[width=0.8\linewidth]{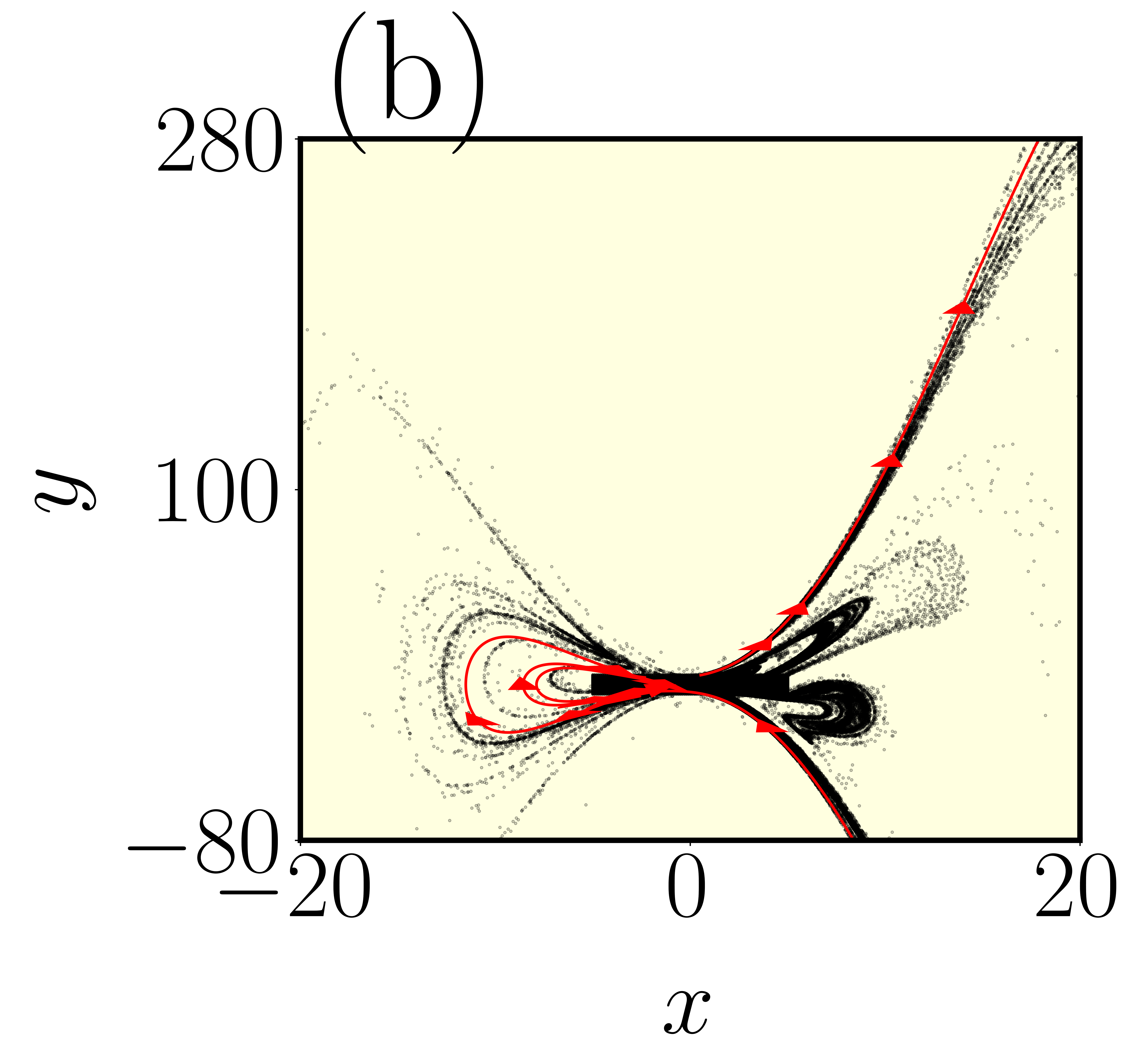}
\caption{(a) Stroboscopic maps revealing chaotic orbits for $f=4.88463$, (b) shows the interior crisis mechanism when chaotic trajectory (red line) meets the detached chaotic orbit leading to long excursion for the forcing parameter value $f=4.88463$.}
\label{fig5}
\end{figure}

As it is well known, the Poincar\'e map provides a crucial tool for categorizing the attractors in the chaotic systems. For example, quasiperiodic orbits have closed point curves and the chaotic attractors have an unlimited number of scattered points on the Poincar\'e surface, and the periodic attractors have a finite number of points on the Poincar\'e section. Also, Poincar\'e maps are often created by intersecting trajectories onto a phase-space hyperplane. Similar to conservative systems, one can employ stroboscopic maps to identify the periodic orbits in non-autonomous system. The dynamical states are sampled at regular time intervals ($T = \frac{2\pi}{\omega}$) for various initial conditions in the phase space in order to create the stroboscopic map. From the stroboscopic map, Fig.\ref{fig5}(a), we see a few orbits that become detached from the bounded chaotic orbits. The trajectory denoted as a red line in Fig.\ref{fig5}(b) runs along and makes a significant phase-space detour when it encounters the detached chaotic orbit in phase space. It does so infrequently and repeatedly, indicating the rarity of extreme events. From the time series plot Fig.\ref{fig3}(b), we observe that EEs occur within certain continuous time intervals. 

\begin{figure}
\centering
\includegraphics[width=0.7\linewidth]{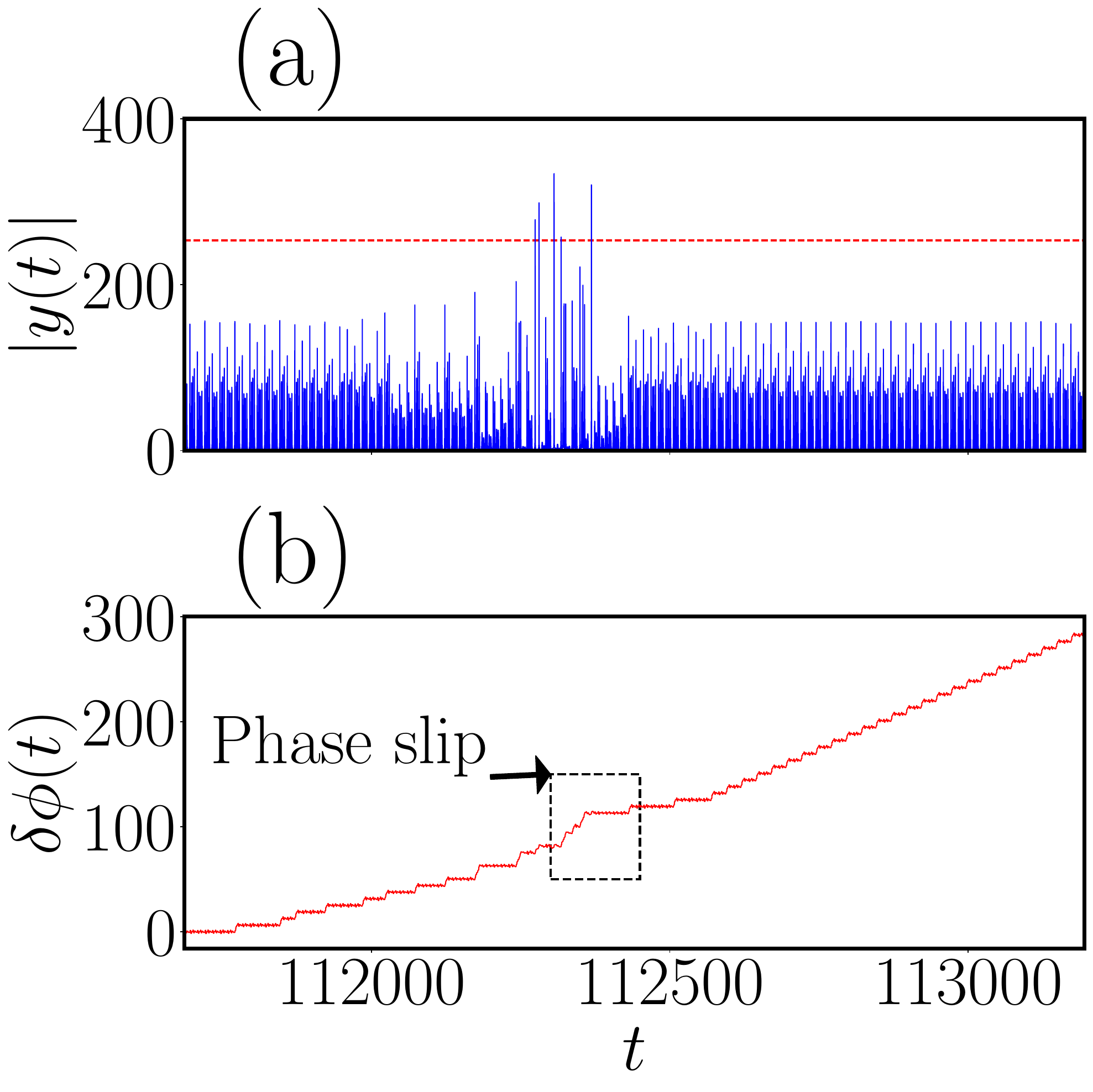}
\caption{{(a) shows the time series and red dashed lines indicate the threshold $H_{s}$. (b) shows the instantaneous phase slips $\delta\phi(t)$  for external forcing parameter $ f=4.88463 $.}}
\label{fig6}
\end{figure}

It is worth noting that a phase slip \cite{park1999phase} occurs between the dynamics of the system and the driving signal when an extreme event occurs. To illustrate this phenomenon, we determine the instantaneous phase $\phi$ of the trajectories of the forced system using a Hilbert transform technique \cite{bhagyaraj2023super}. Subsequently, we proceed to compute the phase difference $\delta\phi$ between the aforementioned trajectories and the external periodic signal $fcos(\omega t)$. Through the examination of the phase disparity between the dynamics of the system and the external periodic signal $f$, it becomes evident that there is substantial variability observed during the manifestation of extreme events (shown in Figs.\ref{fig6}(a) and \ref{fig6}(b)). This observation indicates that the occurrence of phase slip is a primary factor in the manifestation of this phenomenon \cite{kingston2017extreme}. Furthermore, it is observed that the amplitude of the phase difference exhibits a direct correlation with the intensity of the extreme event, providing additional evidence to substantiate the notion that phase slip plays a pivotal role in their manifestation. The aforementioned observations offer significant insights into the fundamental mechanisms that drive severe occurrences, and thus potentially enhancing our ability to foresee and manage them. When intermittency occurs, the phase slip is minimal and rapidly recovers, suggesting that the system can reestablish synchronization with the external signal. 

\begin{figure}
\centering
\includegraphics[width=0.47\linewidth]{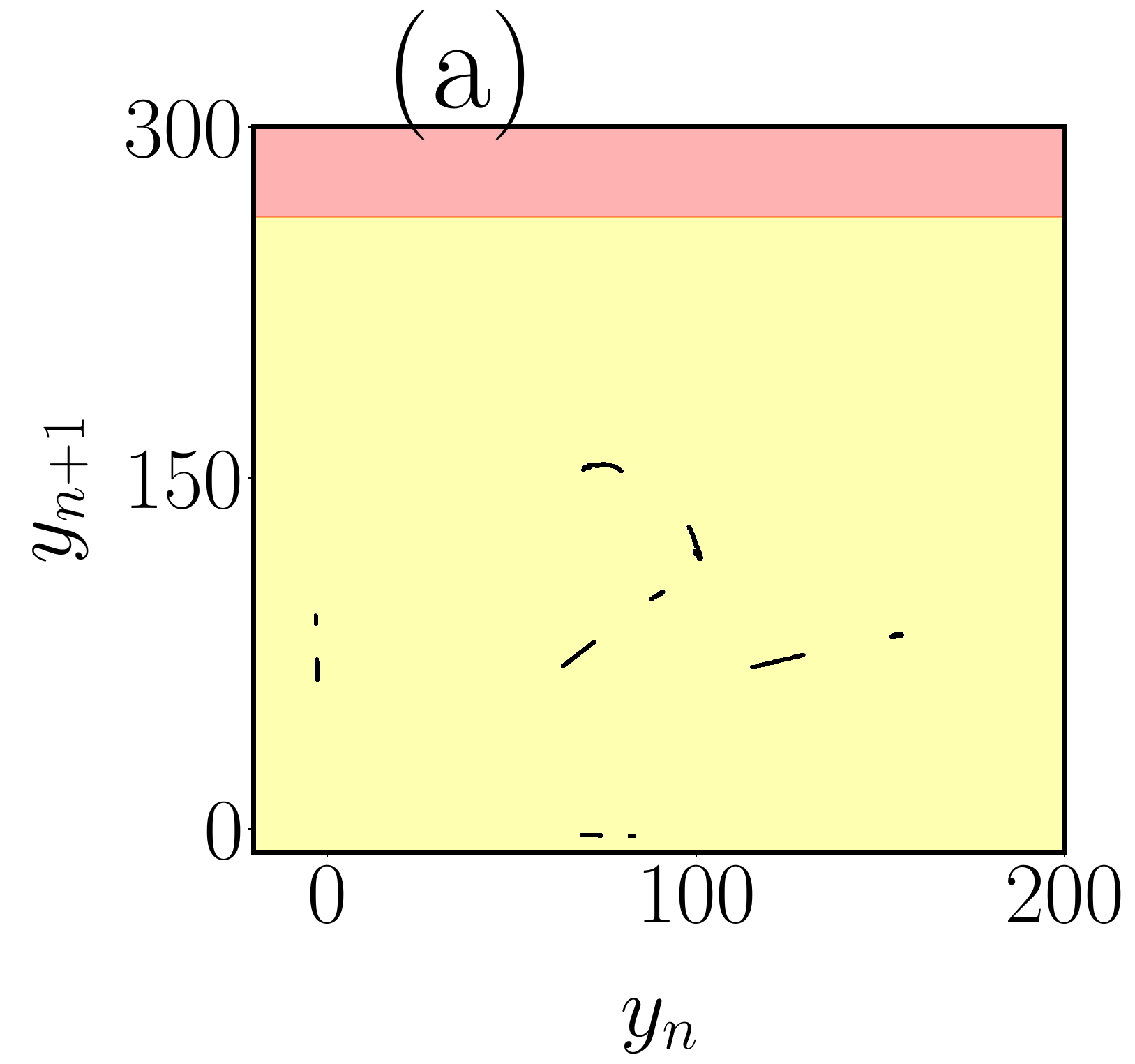}
\includegraphics[width=0.47\linewidth]{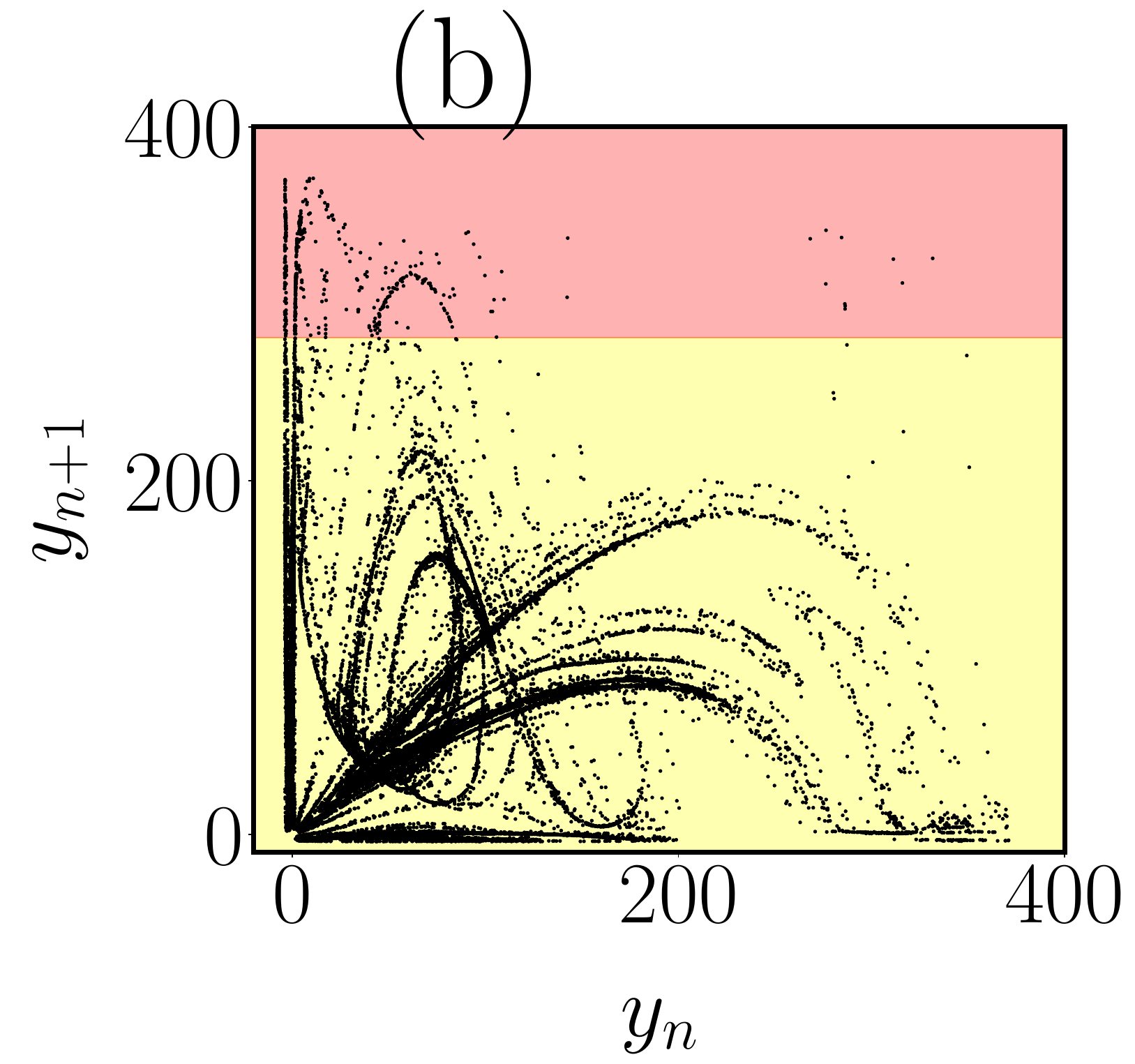}
\includegraphics[width=0.7\linewidth]{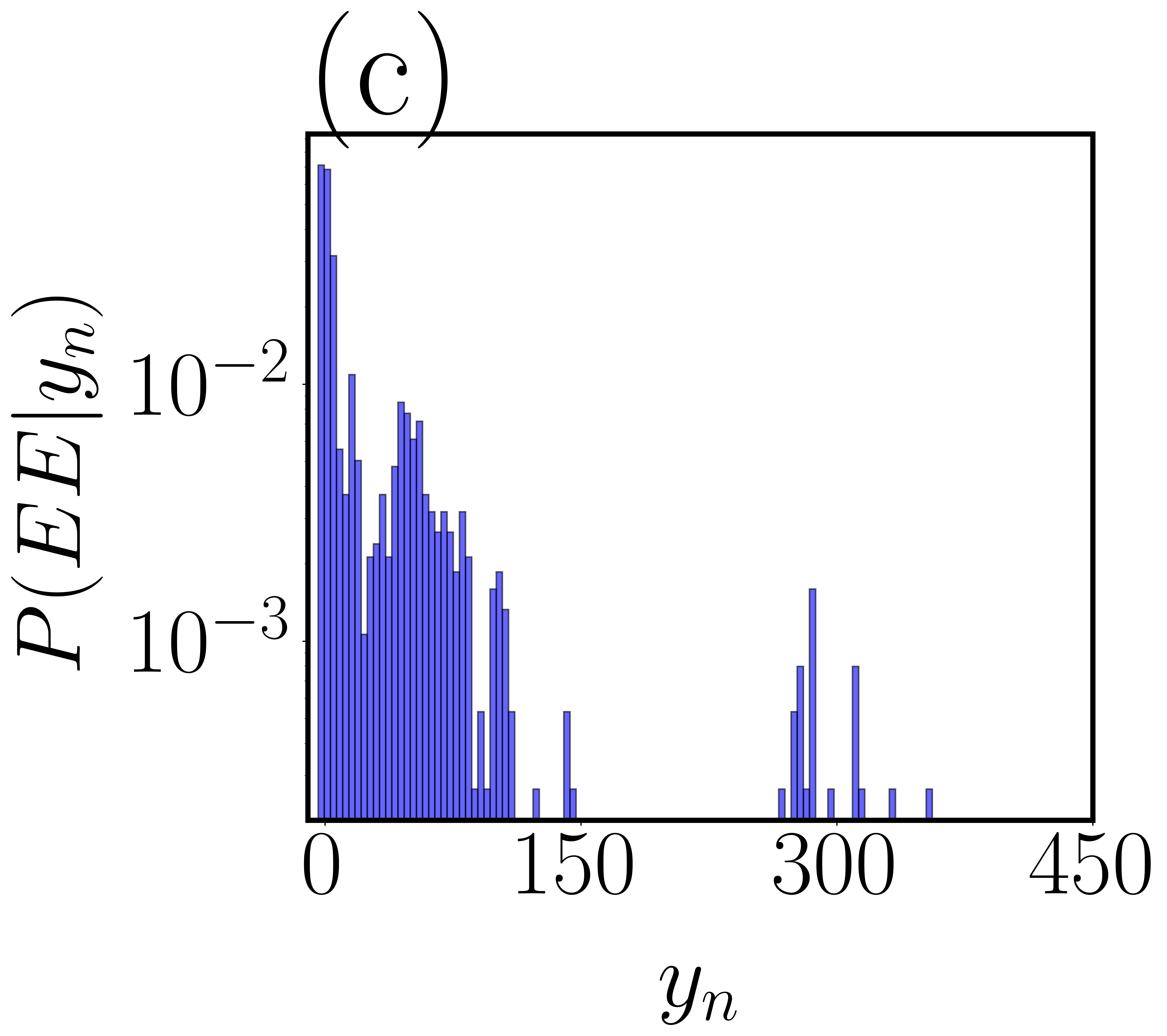}
\caption{{(a) and (b) show the return maps which confirm the occurrence of bounded chaos (dense clusters of black points confined to the yellow region) and extreme events (densely populate the yellow region and sparse points in the red region) respectively at $ f=4.884 \; \& f=4.88463 $, and (c) shows the pre-extreme event distribution for $f=4.88463$.}}
\label{fig7}
\end{figure}
Additionally, we constructed a return map by correlating successive events $y_{n+1}$ with their preceding events $y_{n}$ , enabling a detailed analysis of patterns preceding extreme events. To illustrate, the event space was partitioned into two regions: (1) a bounded event region, shown in yellow, and (2) a region above the extreme event threshold, marked in red. Fig. \ref{fig7}(a) shows dense clusters of black points confined to the yellow region, representing a bounded chaotic regime where events remain below $(y_{n}, y_{n+1}) < 160 $. In Fig.\ref{fig7}(b), under the extreme event (EE) scenario, the black points densely populate the yellow region, indicating bounded chaotic motion with limited amplitude variation between successive events. This is contrasted by sparse points in the red region, signifying an expanded attractor size and underscoring the rarity of EEs.

Furthermore, we observe that extreme post-event occurrences $y_{n+1}$ are most likely when $y_{n}$ values are within the range \( (0.0, 100.0) \), with less frequent occurrences around $y_{n} \approx 300$, as shown in Fig.\ref{fig7}(c).

\section{Statistical characterization of the extreme events}

\begin{figure}
\centering
\includegraphics[width=0.7\linewidth]{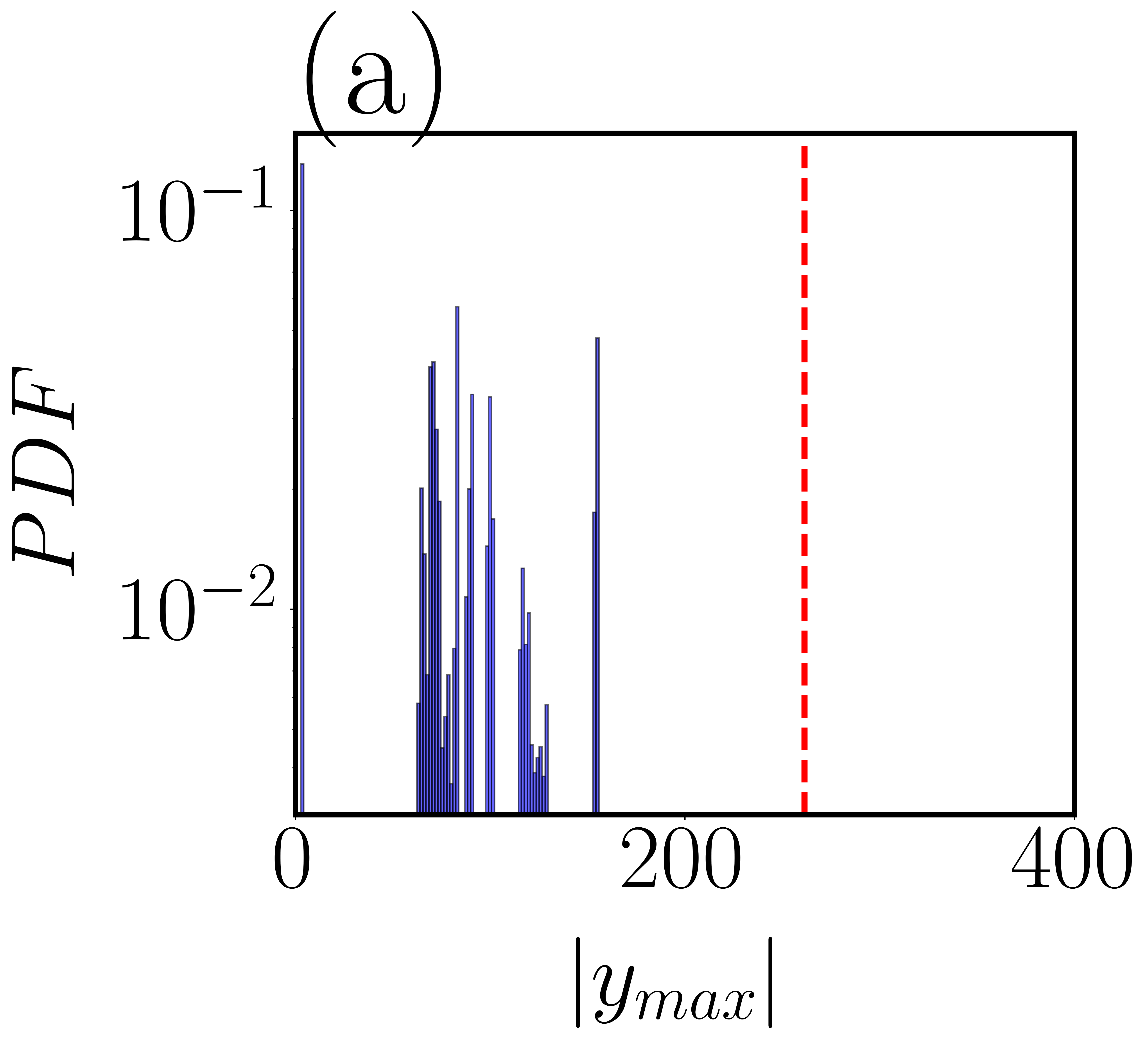}
\includegraphics[width=0.7\linewidth]{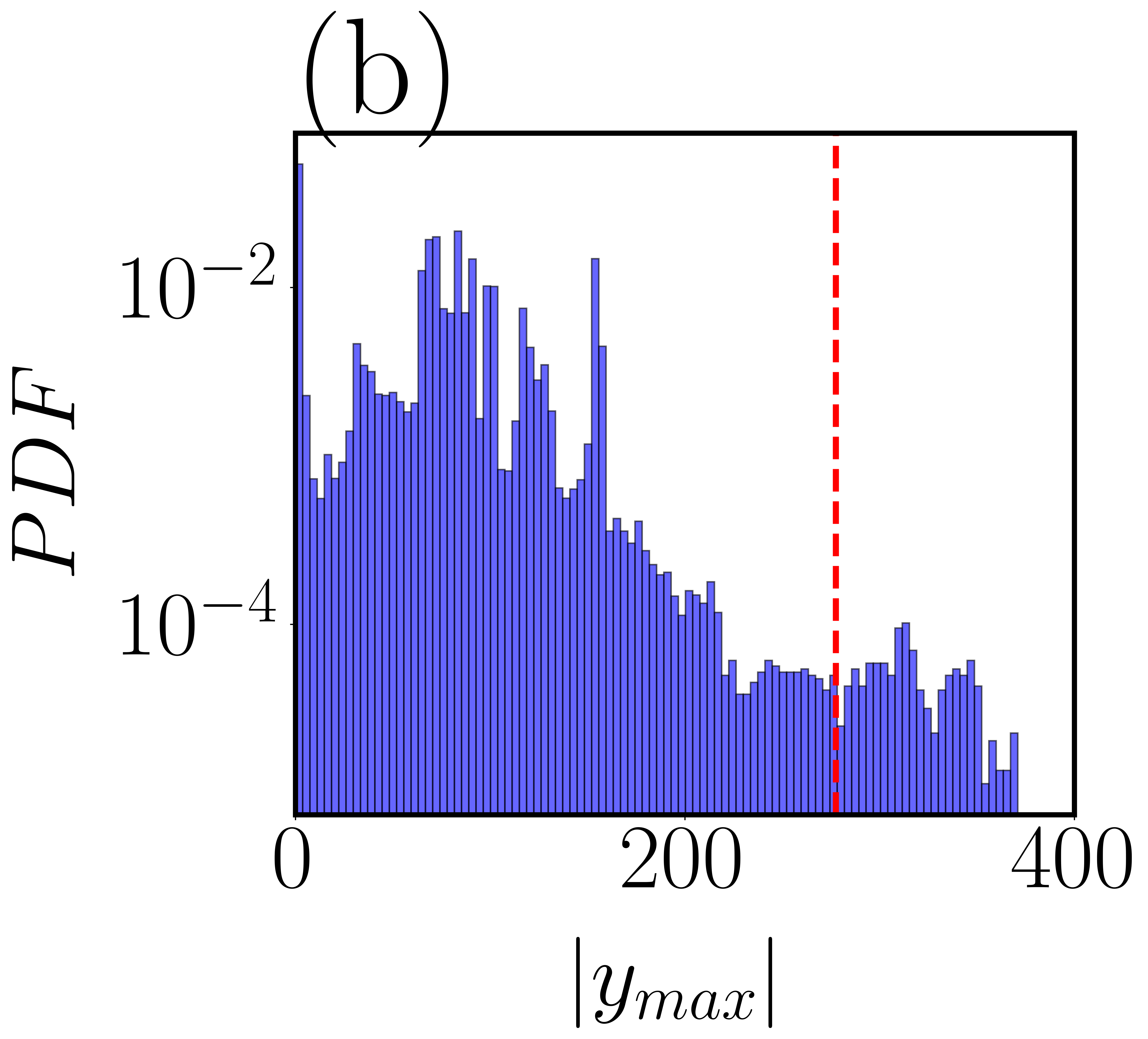}
\caption{{(a) \& (b) show the probability distributions for bounded chaos (PDF distribution ends before the threshold) and extreme events (PDF distributions beyond the threshold) for forcing parameters $f=4.884 \; \& f=4.88463$, respectively. The dashed vertical red color line indicates the threshold values of the observable for categorizing the extreme events.}}
\label{fig8}
\end{figure}

In general, EEs are interpreted as rare and sudden events. One of the important property of EEs, which is widely accepted in the literature is their rare and recurrent occurrences. The rarity of the EEs can be studied through a statistical analysis. In  our statistical characterization, we consider the local maxima  $y_{n}$ of  absolute state variable  $y(t)$. The probability of occurrence of the local maximum is formulated as 
\begin{equation}
probability=\frac{no.~of~particular~y_{n}} {length([y_{n}])} ,
\end{equation}
where $[y_{n}]$ represents the set of all maxima in the time series. The probability distribution function (PDF) for the bounded chaos and the extreme events are shown in Fig.\ref{fig8}(a) and \ref{fig8}(b), respectively. The dashed vertical lines in the PDF distribution plots indicate the threshold values of the observable for categorizing the extreme events. For the case of bounded chaos, the PDF distribution ends before the threshold depicting the zero probability occurrence of EEs. For the case of EEs,  we find the distributions beyond the threshold. Therefore we confirm that the EEs observed in the Higgs oscillator are rare and of less probable occurrences.

\begin{figure}
\centering
\includegraphics[width=0.8\linewidth]{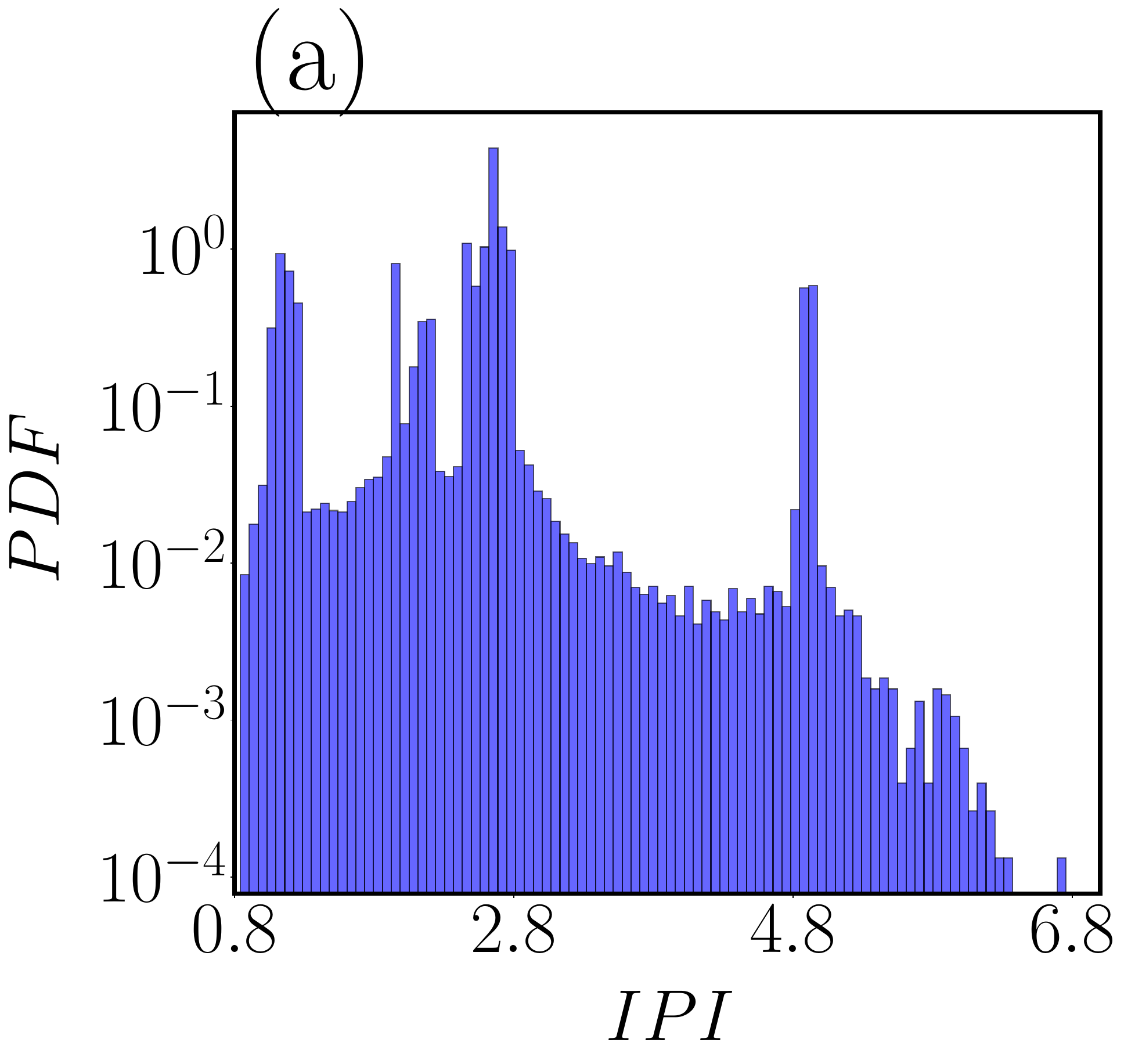}
\includegraphics[width=0.8\linewidth]{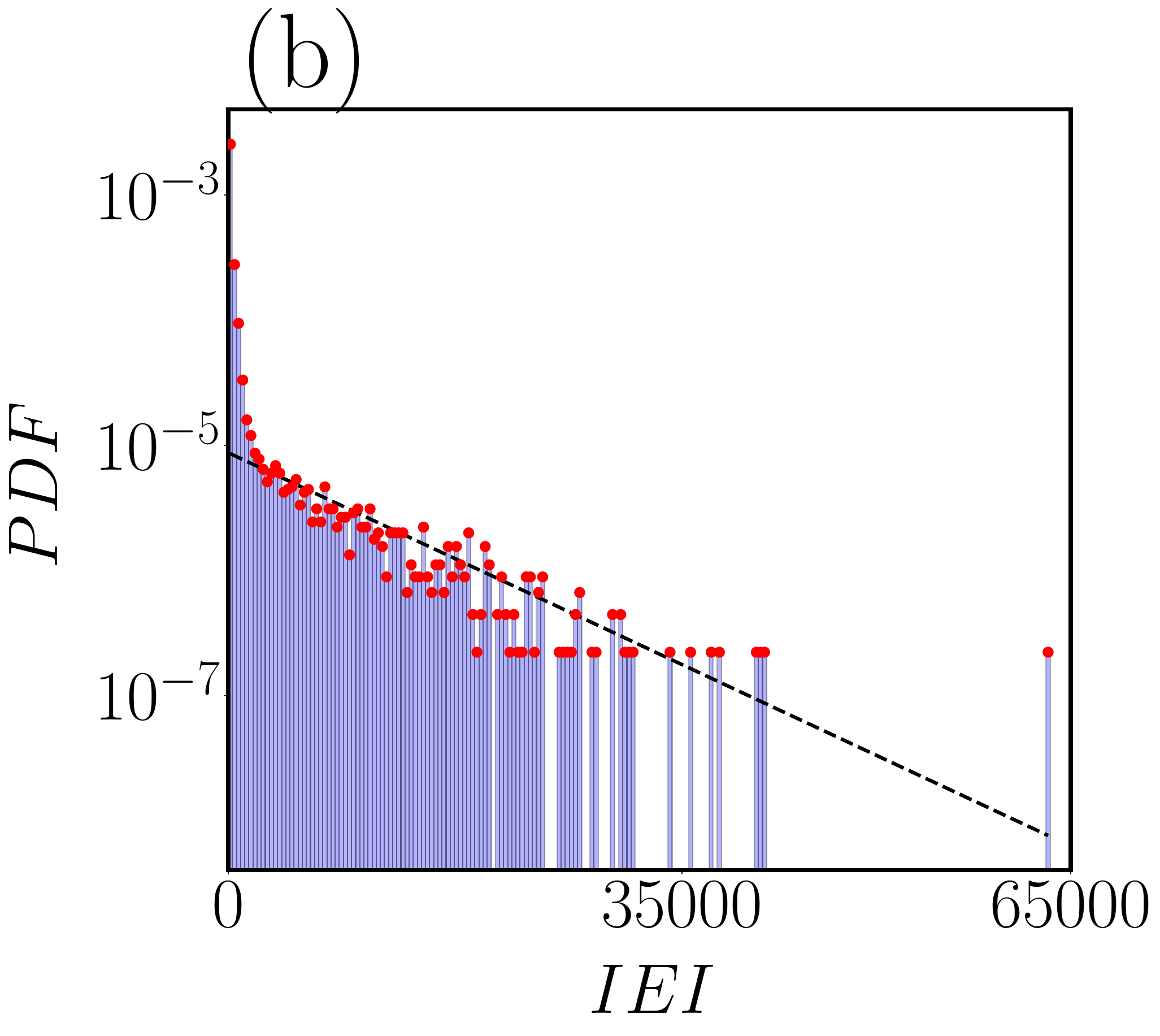}
\caption{{(a) \& (b) show the Inter Peak Interval( IPI ) and  Inter Event Interval( IEI ) distributions for the extreme event, respectively, for $  f=4.88463 $. The red data points in IEI plot depicts the probabilities for corresponding IEI bin centers and the black dashed lines shows the exponential fit corresponding to the exponentially decaying distributions.}}
\label{fig9}
\end{figure}

From many statistical analysis of extreme event intervals, the return interval distributions of EE related time series are best fitted by Weibull distributions \cite{eichner2007statistics,santhanam2008return,blender2008extreme}. In Fig.\ref{fig9}(a), the peak interval distribution of the signal is plotted and the distribution is maximum for the time intervals between $(1.0<IPI<1.5)$ and $(2.0<IPI<2.8)$ and approximately less probable occurrence for the time interval range $(6.0<IPI<6.5)$. The analysis of numerous empirical EE data of the dynamical systems by Santhanam et al. has demonstrated that the inter extreme event interval   distribution is determined by a particular power law distribution, $P(IEI) = Ae^{-B*IEI}$, where IEI is the inter extreme interval, A and B are the shaping parameter of the distributions \cite{santhanam2005long}. Additionally, we illustrate the PDF of the return-time extreme event distribution. And the corresponding shaping parameters $A = 8.719548e-06$ and $B =0.00011138$ are numerically identified using exponential power-law fit. The power law distributions are in close agreement  with the forementioned power-law formulations where we observe the exponentially decaying distributions (see Fig.\ref{fig9}(b)). Although the fit shows the exponentially decaying distributions  and for small IEI, the distributions have some sharp variations from the fit and this variance clearly signifies the clustering of the extreme events (i.e.  successive occurrence of extreme events over some short intervals of time in Higgs oscillator). Thus we have presented substantial evidence from both the PDF and power law distribution to explain the less probable occurrence of EEs and recurrence over time.

\section{Forecasting Extreme Events}

Sudden appearance of large amplitude oscillations has a huge and disastrous impact on nature ranging from earthquakes, tsunamis, and stock market crash to epileptic seizures in the brain, etc \cite{chowdhury2022extreme}. Such occurrence of rare and unexpected dynamics can be realized by identifying a region of instability in phase space. All such realizations help in mitigating the severity of the events;  another one of the most trending approaches is by utilizing the artificial intelligence (AI) and machine learning (ML) algorithms to forecast such events.

Previous studies and many experimentations with forecasting sequential data shows that the RNN algorithm performs best among the other ML algorithms like convolutional neural networks (CNN) and multilayer perceptron (MLP). Although RNN has a huge advantage of predicting long-term correlated data, it suffers from vanishing gradient problems during updation of neuronal weights and biases. Such difficulties are replaced by a new RNN-based architecture known as LSTM (Long Short-Term Memory) \cite{lstm}. Inside the LSTM architecture three main gates control the flow of information: the input gate, forget gate, and output gate. These gates regulate what information is passed on, updated, or discarded, helping the LSTM to remember or forget data over time \cite{lstm}.

\begin{figure}[!h]
\centering
\includegraphics[width=0.8\linewidth]{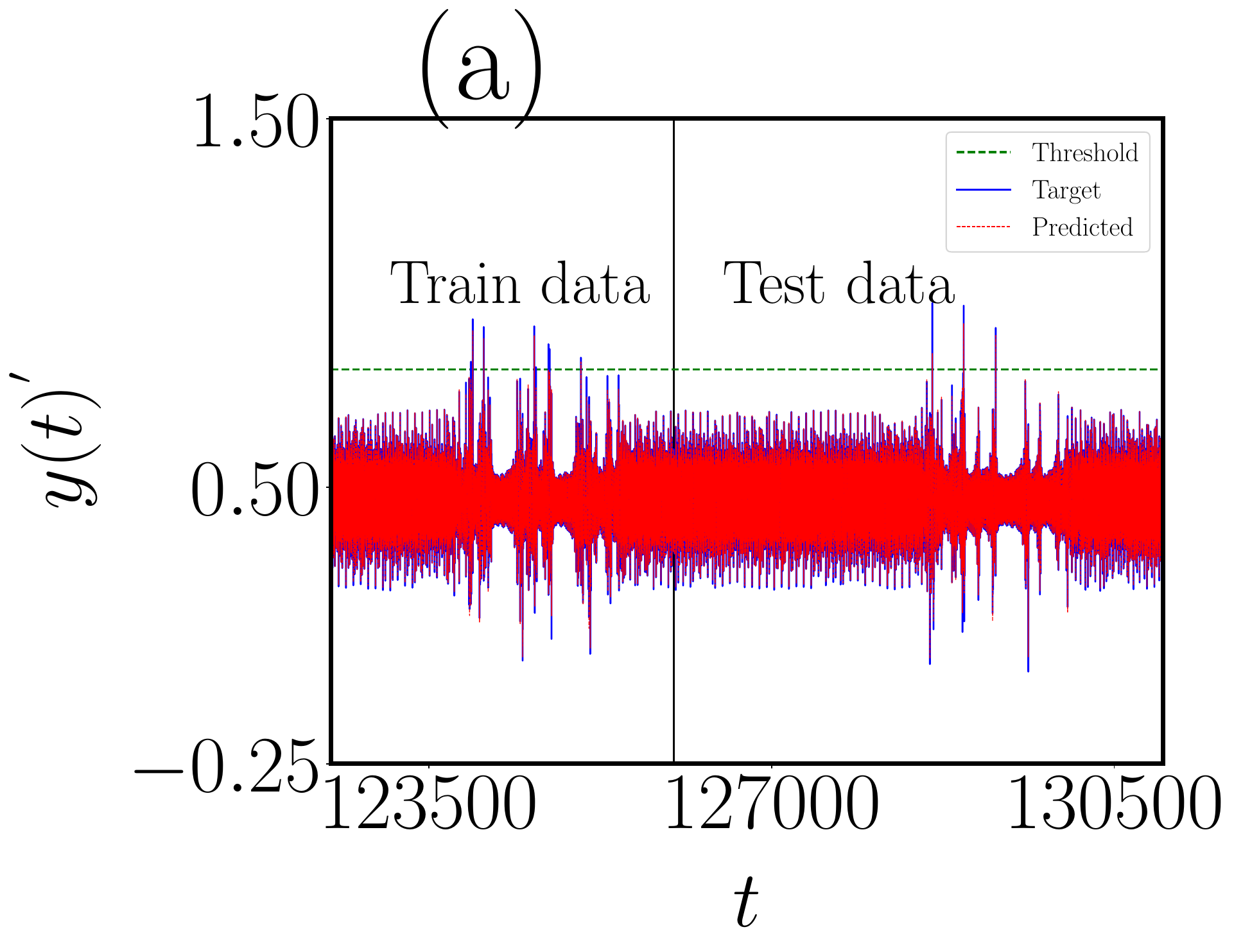}
\includegraphics[width=0.8\linewidth]{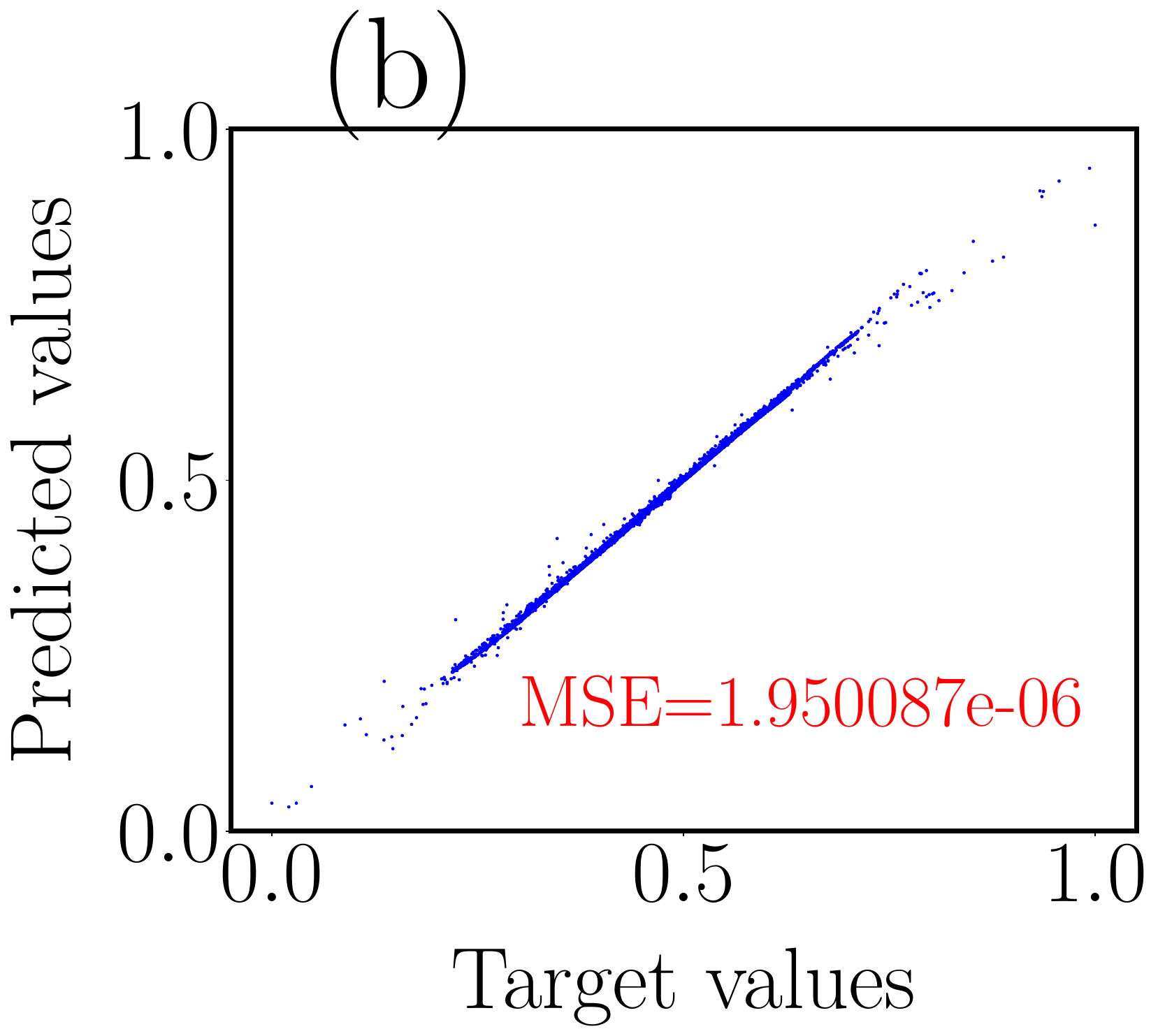}
\caption{(a) shows the EEs associated time series prediction for control parameter $  f=4.88463 $  using LSTM, blue line -target data, red dashed line - predicted data, green dashed line - qualifying threshold for EEs, black line - To separate training and testing phase, (b) show the corresponding correlation plot for target vs predicted values.}
\label{fig10}
\end{figure}

In recent years, parameter-aware reservoir computing has emerged as a successful method for replicating bifurcation scenarios in dynamical systems \cite{xiao2021predicting}. However, this work focuses on forecasting the extreme events. In this section, we utilize LSTM to predict the extreme events (EEs) exhibited by the damped-driven Higgs oscillator. For training and forecasting the extreme events, we use 85,002 data points so as to sufficiently cover the region of extreme events, which span specific time intervals shown in Fig.~\ref{fig10}(a) from our previous studies at the parameter value $f=4.88463$ to demonstrate the forecasting process. Before forecasting the EE time series data, we first normalize the data using Min-Max normalization and make the data distributed between 0 and 1.0, 
\begin{equation}
    y' = \frac{y - \min(y)}{\max(y) - \min(y)} .
\end{equation}
After normalizing the time series data using the above relation, 
we structure the LSTM model for forecasting the sequence data. In our present work, we choose two-layer LSTM with 40 LSTM neuron cells each and a single output dense layer. The number of LSTM layers,  and the number of neurons in each of the layers are the hyper parameters that can be optimized by a series of trials. We use the python package KERAS to construct the LSTM model. Having constructed the model, we must organize the input and target data for training the LSTM model as shown in Table \ref{table:1}. The input data can be a past data ($y^{'}_{n-1}$) or can be the past set of data $[y^{'}_{1},y^{'}_{2},y^{'}_{3},..y^{'}_{n-1}]$ and the target should be the future data $y^{'}_{n}$. In our work, we consider past sets of data with sliding window size of 20 data as shown in Table \ref{table:1}. When we feed the input as past sets of the data $[y^{'}_{1},y^{'}_{2},y^{'}_{3},..y^{'}_{n-1}]$ into the LSTM model we expect the output data to be $y^{'}_{n}$. Initially, the LSTM model gives the output $y^{'*}_{n}$ which is a deviated value from the target value, and the error between them is calculated using the mean square error (MSE) relation,

 \begin{equation}
MSE = \sum_{n=1}^{N_{sample}}\frac{1}{N_{sample}}(y^{'*}(n) - y^{'}(n))^{2}.
\end{equation}

\begin{table}[h]
  \centering
  \begin{tabular}{|p{3cm}|p{2cm}|p{2cm}|}
    \hline
    \multicolumn{3}{|c|}{Data structuring for training the LSTM model} \\
    \hline
    Input data & Target data & Predicted Output\\
    \hline
    $y^{'}_{1},y^{'}_{2},y^{'}_{3},...y^{'}_{20}$ & $y^{'}_{21}$ & $y^{'*}_{21}$  \\ & & \\
    \hline
    $y^{'}_{2},y^{'}_{3},y^{'}_{4},...y^{'}_{21}$ & $y^{'}_{22}$ & $y^{'*}_{22}$ \\ &  &\\
    \hline
    .... & . & .\\
    \hline
    .... & . & .\\
    \hline
    .... & . & . \\
    \hline
    .... & . & .\\
    \hline
    $y^{'}_{34980},y^{'}_{34981},...y^{'}_{34999}$ & $y^{'}_{35000}$ & $y^{'*}_{35000}$  \\ &  &\\
    \hline
  \end{tabular}
  \caption{Data preparation table for training the LSTM model.}
  \label{table:1}
\end{table}

The error deviations between the predicted and target values decrease as the values of weights and biases of the input, update, and forget gate are updated for every training iteration. We increase the number of training iterations so that the error can be reduced at each step until reaching the saturation point where the error cannot be further reduced and it oscillates in a small range. Finally, after the whole iteration, we obtain the updated weights and biases in the gates of every LSTM cell in the layers. In our work, we have taken 35000 data for training the LSTM model and for optimizing the model, and we use Adam optimizer with 250 training iterations which is sufficient enough to reach the error saturation.

After accomplishing the above steps, the LSTM model is ready to enter into the testing phase. The model is tested for remaining 49982 data points where this testing data (blue solid line) comprises of intermittent high amplitude extreme event oscillations. In Fig.\ref{fig10}(a) we can see the predicted output (red line) is able to nearly 
capture the sudden intermittent extreme event transitions, where these EEs transitions are confirmed by fixing the EE threshold for normalized time series ($y(t)^{'}$) using the relations $ H_{s}=<y^{'}_{n}>+N\sigma(y^{'}_{n})$ and the threshold for normalized time series is found to be $ H_{s}=0.8193$. Also the correlation plot is illustrated in Fig.\ref{fig10}(b) to show the closeness of the target and the predicted values, and the MSE between them is found to be $1.950087e^{-6}$. Thus we are able to forecast the critical EEs transitions from the bounded chaotic time series data using the LSTM model. 

\section{Conclusion}
 
In summary, we have studied the dynamics of the one-dimensional Higgs oscillator in the presence of damping and an external force to explore various dynamical behaviors across different parameters, particularly the large amplitude behavior and interior crises. The large amplitude behavior is characterized by a probability distribution function that confirms extreme events. Additionally, we investigated the mechanism underlying the emergence of these extreme events. We further characterized the large amplitude behavior using stroboscopic maps, phase slips, Inter-Peak Intervals (IPI), and Inter-Event Intervals (IEI) and confirmed the extreme events (EEs). Finally, we have demonstrated the prediction of  EEs in the damped, driven Higgs oscillator  using  LSTM model. On the whole, this study will add critical information for understanding the bifurcation process and the mechanism of generation of EEs involved in non-polynomial oscillators.  

\section*{AUTHOR’S CONTRIBUTIONS}		
All authors contributed equally.

\section*{Acknowledgments}
The author W.A.M acknowledges DST- INSPIRE, Govt. of India for the award of Junior Research Fellowship (DST INSPIRE – JRF IF210722/DST Inspire Fellowship/2022), A.V acknowledges the DST-FIST for funding research projects via Grant No.SR/FST/College-2018-372(C). M.L wishes to thank the Department of Science and Technology for the award of a DST-SERB National Science Chair under Grant No.NSC/2020/00029 in which  M.S is supported by Research Associateship. V. C. would like to thank SRM TRP Engineering College, India, for their financial support, vide number SRM/TRP/RI/005.

\section*{DATA AVAILABILITY}
The data supporting the findings of this study are available from the authors on reasonable request.

\section*{REFERENCES}


\begin{thebibliography}{76}%
	\makeatletter
	\providecommand \@ifxundefined [1]{%
		\@ifx{#1\undefined}
	}%
	\providecommand \@ifnum [1]{%
		\ifnum #1\expandafter \@firstoftwo
		\else \expandafter \@secondoftwo
		\fi
	}%
	\providecommand \@ifx [1]{%
		\ifx #1\expandafter \@firstoftwo
		\else \expandafter \@secondoftwo
		\fi
	}%
	\providecommand \natexlab [1]{#1}%
	\providecommand \enquote  [1]{``#1''}%
	\providecommand \bibnamefont  [1]{#1}%
	\providecommand \bibfnamefont [1]{#1}%
	\providecommand \citenamefont [1]{#1}%
	\providecommand \href@noop [0]{\@secondoftwo}%
	\providecommand \href [0]{\begingroup \@sanitize@url \@href}%
	\providecommand \@href[1]{\@@startlink{#1}\@@href}%
	\providecommand \@@href[1]{\endgroup#1\@@endlink}%
	\providecommand \@sanitize@url [0]{\catcode `\\12\catcode `\$12\catcode
		`\&12\catcode `\#12\catcode `\^12\catcode `\_12\catcode `\%12\relax}%
	\providecommand \@@startlink[1]{}%
	\providecommand \@@endlink[0]{}%
	\providecommand \url  [0]{\begingroup\@sanitize@url \@url }%
	\providecommand \@url [1]{\endgroup\@href {#1}{\urlprefix }}%
	\providecommand \urlprefix  [0]{URL }%
	\providecommand \Eprint [0]{\href }%
	\providecommand \doibase [0]{http://dx.doi.org/}%
	\providecommand \selectlanguage [0]{\@gobble}%
	\providecommand \bibinfo  [0]{\@secondoftwo}%
	\providecommand \bibfield  [0]{\@secondoftwo}%
	\providecommand \translation [1]{[#1]}%
	\providecommand \BibitemOpen [0]{}%
	\providecommand \bibitemStop [0]{}%
	\providecommand \bibitemNoStop [0]{.\EOS\space}%
	\providecommand \EOS [0]{\spacefactor3000\relax}%
	\providecommand \BibitemShut  [1]{\csname bibitem#1\endcsname}%
	\let\auto@bib@innerbib\@empty
	\bibitem [{\citenamefont {Mathews}\ and\ \citenamefont
		{Lakshmanan}(1974)}]{mathews1974unique}%
	\BibitemOpen
	\bibfield  {author} {\bibinfo {author} {\bibfnamefont {P.~M.}\ \bibnamefont
			{Mathews}}\ and\ \bibinfo {author} {\bibfnamefont {M.}~\bibnamefont
			{Lakshmanan}},\ }\bibfield  {title} {\enquote {\bibinfo {title} {On a unique
				nonlinear oscillator},}\ }\href@noop {} {\bibfield  {journal} {\bibinfo
			{journal} {Q. Appl. Math.}\ }\textbf {\bibinfo {volume} {32}},\ \bibinfo
		{pages} {215--218} (\bibinfo {year} {1974})}\BibitemShut {NoStop}%
	\bibitem [{\citenamefont {Mathews}\ and\ \citenamefont
		{Lakshmanan}(1975)}]{mathews1975quantum}%
	\BibitemOpen
	\bibfield  {author} {\bibinfo {author} {\bibfnamefont {P.~M.}\ \bibnamefont
			{Mathews}}\ and\ \bibinfo {author} {\bibfnamefont {M.}~\bibnamefont
			{Lakshmanan}},\ }\bibfield  {title} {\enquote {\bibinfo {title} {A
				quantum-mechanically solvable nonpolynomial lagrangian with
				velocity-dependent interaction},}\ }\href@noop {} {\bibfield  {journal}
		{\bibinfo  {journal} {Nuovo Cim. A}\ }\textbf {\bibinfo {volume} {26}},\
		\bibinfo {pages} {299--316} (\bibinfo {year} {1975})}\BibitemShut {NoStop}%
	\bibitem [{\citenamefont {Higgs}(1979)}]{higgs1979dynamical}%
	\BibitemOpen
	\bibfield  {author} {\bibinfo {author} {\bibfnamefont {P.~W.}\ \bibnamefont
			{Higgs}},\ }\bibfield  {title} {\enquote {\bibinfo {title} {Dynamical
				symmetries in a spherical geometry. i},}\ }\href@noop {} {\bibfield
		{journal} {\bibinfo  {journal} {J. Phys. A: Math. Gen.}\ }\textbf {\bibinfo
			{volume} {12}},\ \bibinfo {pages} {309} (\bibinfo {year} {1979})}\BibitemShut
	{NoStop}%
	\bibitem [{\citenamefont {Lakshmanan}\ and\ \citenamefont
		{Eswaran}(1975)}]{lakshmanan1975quantum}%
	\BibitemOpen
	\bibfield  {author} {\bibinfo {author} {\bibfnamefont {M.}~\bibnamefont
			{Lakshmanan}}\ and\ \bibinfo {author} {\bibfnamefont {K.}~\bibnamefont
			{Eswaran}},\ }\bibfield  {title} {\enquote {\bibinfo {title} {Quantum
				dynamics of a solvable nonlinear chiral model},}\ }\href@noop {} {\bibfield
		{journal} {\bibinfo  {journal} {J. Phys. A: Math. Gen.}\ }\textbf {\bibinfo
			{volume} {8}},\ \bibinfo {pages} {1658} (\bibinfo {year} {1975})}\BibitemShut
	{NoStop}%
	\bibitem [{\citenamefont {Ra{\~n}ada}\ and\ \citenamefont
		{Santander}(2002)}]{ranada2002harmonic}%
	\BibitemOpen
	\bibfield  {author} {\bibinfo {author} {\bibfnamefont {M.~F.}\ \bibnamefont
			{Ra{\~n}ada}}\ and\ \bibinfo {author} {\bibfnamefont {M.}~\bibnamefont
			{Santander}},\ }\bibfield  {title} {\enquote {\bibinfo {title} {On harmonic
				oscillators on the two-dimensional sphere s 2 and the hyperbolic plane h
				2},}\ }\href@noop {} {\bibfield  {journal} {\bibinfo  {journal} {J. Math.
				Phys.}\ }\textbf {\bibinfo {volume} {43}},\ \bibinfo {pages} {431--451}
		(\bibinfo {year} {2002})}\BibitemShut {NoStop}%
	\bibitem [{\citenamefont {Lakshmanan}\ and\ \citenamefont
		{Chandrasekar}(2013)}]{lakshmanan2013generating}%
	\BibitemOpen
	\bibfield  {author} {\bibinfo {author} {\bibfnamefont {M.}~\bibnamefont
			{Lakshmanan}}\ and\ \bibinfo {author} {\bibfnamefont {V.~K.}\ \bibnamefont
			{Chandrasekar}},\ }\bibfield  {title} {\enquote {\bibinfo {title} {Generating
				finite dimensional integrable nonlinear dynamical systems},}\ }\href@noop {}
	{\bibfield  {journal} {\bibinfo  {journal} {Eur. Phys. J. Spec. Top}\
		}\textbf {\bibinfo {volume} {222}},\ \bibinfo {pages} {665--688} (\bibinfo
		{year} {2013})}\BibitemShut {NoStop}%
	\bibitem [{\citenamefont {Quesne}(2018)}]{quesne2018deformed}%
	\BibitemOpen
	\bibfield  {author} {\bibinfo {author} {\bibfnamefont {C.}~\bibnamefont
			{Quesne}},\ }\bibfield  {title} {\enquote {\bibinfo {title} {Deformed shape
				invariance symmetry and potentials in curved space with two known
				eigenstates},}\ }\href@noop {} {\bibfield  {journal} {\bibinfo  {journal} {J.
				Math. Phys.}\ }\textbf {\bibinfo {volume} {59}},\ \bibinfo {pages} {042104}
		(\bibinfo {year} {2018})}\BibitemShut {NoStop}%
	\bibitem [{\citenamefont {Ballesteros}\ \emph {et~al.}(2008)\citenamefont
		{Ballesteros}, \citenamefont {Enciso}, \citenamefont {Herranz},\ and\
		\citenamefont {Ragnisco}}]{ballesteros2008superintegrable}%
	\BibitemOpen
	\bibfield  {author} {\bibinfo {author} {\bibfnamefont {A.}~\bibnamefont
			{Ballesteros}}, \bibinfo {author} {\bibfnamefont {A.}~\bibnamefont {Enciso}},
		\bibinfo {author} {\bibfnamefont {F.~J.}\ \bibnamefont {Herranz}}, \ and\
		\bibinfo {author} {\bibfnamefont {O.}~\bibnamefont {Ragnisco}},\ }\bibfield
	{title} {\enquote {\bibinfo {title} {Superintegrable anharmonic oscillators
				on n-dimensional curved spaces},}\ }\href@noop {} {\bibfield  {journal}
		{\bibinfo  {journal} {J. Nonlinear Math. Phys.}\ }\textbf {\bibinfo {volume}
			{15}},\ \bibinfo {pages} {43--52} (\bibinfo {year} {2008})}\BibitemShut
	{NoStop}%
	\bibitem [{\citenamefont {Chithiika~Ruby}\ and\ \citenamefont
		{Lakshmanan}(2021)}]{ruby2021classical}%
	\BibitemOpen
	\bibfield  {author} {\bibinfo {author} {\bibfnamefont {V.}~\bibnamefont
			{Chithiika~Ruby}}\ and\ \bibinfo {author} {\bibfnamefont {M.}~\bibnamefont
			{Lakshmanan}},\ }\bibfield  {title} {\enquote {\bibinfo {title} {On the
				classical and quantum dynamics of a class of nonpolynomial oscillators},}\
	}\href@noop {} {\bibfield  {journal} {\bibinfo  {journal} {J. Phys. A: Math.
				Theor}\ }\textbf {\bibinfo {volume} {54}},\ \bibinfo {pages} {385301}
		(\bibinfo {year} {2021})}\BibitemShut {NoStop}%
	\bibitem [{\citenamefont {Chithiika~Ruby}\ and\ \citenamefont
		{Lakshmanan}(2024)}]{V_2024}%
	\BibitemOpen
	\bibfield  {author} {\bibinfo {author} {\bibfnamefont {V.}~\bibnamefont
			{Chithiika~Ruby}}\ and\ \bibinfo {author} {\bibfnamefont {M.}~\bibnamefont
			{Lakshmanan}},\ }\bibfield  {title} {\enquote {\bibinfo {title} {Liénard
				type nonlinear oscillators and quantum solvability},}\ }\href@noop {}
	{\bibfield  {journal} {\bibinfo  {journal} {Phys. Scr.}\ }\textbf {\bibinfo
			{volume} {99}},\ \bibinfo {pages} {062004} (\bibinfo {year}
		{2024})}\BibitemShut {NoStop}%
	\bibitem [{\citenamefont {Hakobyan}, \citenamefont {Nersessian},\ and\
		\citenamefont {Yeghikyan}(2009)}]{hakobyan2009cuboctahedric}%
	\BibitemOpen
	\bibfield  {author} {\bibinfo {author} {\bibfnamefont {T.}~\bibnamefont
			{Hakobyan}}, \bibinfo {author} {\bibfnamefont {A.}~\bibnamefont
			{Nersessian}}, \ and\ \bibinfo {author} {\bibfnamefont {V.}~\bibnamefont
			{Yeghikyan}},\ }\bibfield  {title} {\enquote {\bibinfo {title} {The
				cuboctahedric higgs oscillator from the rational calogero model},}\
	}\href@noop {} {\bibfield  {journal} {\bibinfo  {journal} {J. Phys. A: Math.
				Theor}\ }\textbf {\bibinfo {volume} {42}},\ \bibinfo {pages} {205206}
		(\bibinfo {year} {2009})}\BibitemShut {NoStop}%
	\bibitem [{\citenamefont {Mohammadi}, \citenamefont {Aghaei},\ and\
		\citenamefont {Chenaghlou}(2016)}]{mohammadi2016dirac}%
	\BibitemOpen
	\bibfield  {author} {\bibinfo {author} {\bibfnamefont {V.}~\bibnamefont
			{Mohammadi}}, \bibinfo {author} {\bibfnamefont {S.}~\bibnamefont {Aghaei}}, \
		and\ \bibinfo {author} {\bibfnamefont {A.}~\bibnamefont {Chenaghlou}},\
	}\bibfield  {title} {\enquote {\bibinfo {title} {Dirac equation in presence
				of the hartmann and higgs oscillator superintegrable potentials with the spin
				and pseudospin symmetries},}\ }\href@noop {} {\bibfield  {journal} {\bibinfo
			{journal} {Int. J. Mod. Phys. A}\ }\textbf {\bibinfo {volume} {31}},\
		\bibinfo {pages} {1650190} (\bibinfo {year} {2016})}\BibitemShut {NoStop}%
	\bibitem [{\citenamefont {Carinena}\ \emph {et~al.}(2004)\citenamefont
		{Carinena}, \citenamefont {Ranada}, \citenamefont {Santander},\ and\
		\citenamefont {Senthilvelan}}]{carinena2004non}%
	\BibitemOpen
	\bibfield  {author} {\bibinfo {author} {\bibfnamefont {J.~F.}\ \bibnamefont
			{Carinena}}, \bibinfo {author} {\bibfnamefont {M.~F.}\ \bibnamefont
			{Ranada}}, \bibinfo {author} {\bibfnamefont {M.}~\bibnamefont {Santander}}, \
		and\ \bibinfo {author} {\bibfnamefont {M.}~\bibnamefont {Senthilvelan}},\
	}\bibfield  {title} {\enquote {\bibinfo {title} {A non-linear oscillator with
				quasi-harmonic behaviour: two-and n-dimensional oscillators},}\ }\href@noop
	{} {\bibfield  {journal} {\bibinfo  {journal} {Nonlinearity}\ }\textbf
		{\bibinfo {volume} {17}},\ \bibinfo {pages} {1941} (\bibinfo {year}
		{2004})}\BibitemShut {NoStop}%
	\bibitem [{\citenamefont {Mahdifar}, \citenamefont {Roknizadeh},\ and\
		\citenamefont {Naderi}(2006)}]{mahdifar2006geometric}%
	\BibitemOpen
	\bibfield  {author} {\bibinfo {author} {\bibfnamefont {A.}~\bibnamefont
			{Mahdifar}}, \bibinfo {author} {\bibfnamefont {R.}~\bibnamefont
			{Roknizadeh}}, \ and\ \bibinfo {author} {\bibfnamefont {M.}~\bibnamefont
			{Naderi}},\ }\bibfield  {title} {\enquote {\bibinfo {title} {Geometric
				approach to nonlinear coherent states using the higgs model for harmonic
				oscillator},}\ }\href@noop {} {\bibfield  {journal} {\bibinfo  {journal} {J.
				Phys. A Math. Gen.}\ }\textbf {\bibinfo {volume} {39}},\ \bibinfo {pages}
		{7003} (\bibinfo {year} {2006})}\BibitemShut {NoStop}%
	\bibitem [{\citenamefont {Evnin}\ and\ \citenamefont
		{Nivesvivat}(2016)}]{evnin2016ads}%
	\BibitemOpen
	\bibfield  {author} {\bibinfo {author} {\bibfnamefont {O.}~\bibnamefont
			{Evnin}}\ and\ \bibinfo {author} {\bibfnamefont {R.}~\bibnamefont
			{Nivesvivat}},\ }\bibfield  {title} {\enquote {\bibinfo {title} {Ads
				perturbations, isometries, selection rules and the higgs oscillator},}\
	}\href@noop {} {\bibfield  {journal} {\bibinfo  {journal} {JHEP}\ }\textbf
		{\bibinfo {volume} {2016}},\ \bibinfo {pages} {1--25} (\bibinfo {year}
		{2016})}\BibitemShut {NoStop}%
	\bibitem [{\citenamefont {Venkatesan}\ and\ \citenamefont
		{Lakshmanan}(1997)}]{venkatesan1997nonlinear}%
	\BibitemOpen
	\bibfield  {author} {\bibinfo {author} {\bibfnamefont {A.}~\bibnamefont
			{Venkatesan}}\ and\ \bibinfo {author} {\bibfnamefont {M.}~\bibnamefont
			{Lakshmanan}},\ }\bibfield  {title} {\enquote {\bibinfo {title} {Nonlinear
				dynamics of damped and driven velocity-dependent systems},}\ }\href@noop {}
	{\bibfield  {journal} {\bibinfo  {journal} {Phys. Rev. E}\ }\textbf {\bibinfo
			{volume} {55}},\ \bibinfo {pages} {5134} (\bibinfo {year}
		{1997})}\BibitemShut {NoStop}%
	\bibitem [{\citenamefont {Ott}(2002)}]{ott2002chaos}%
	\BibitemOpen
	\bibfield  {author} {\bibinfo {author} {\bibfnamefont {E.}~\bibnamefont
			{Ott}},\ }\href@noop {} {\emph {\bibinfo {title} {Chaos in dynamical
				systems}}}\ (\bibinfo  {publisher} {Cambridge university press},\ \bibinfo
	{year} {2002})\BibitemShut {NoStop}%
	\bibitem [{\citenamefont {Grebogi}, \citenamefont {Ott},\ and\ \citenamefont
		{Yorke}(1983)}]{grebogi1983crises}%
	\BibitemOpen
	\bibfield  {author} {\bibinfo {author} {\bibfnamefont {C.}~\bibnamefont
			{Grebogi}}, \bibinfo {author} {\bibfnamefont {E.}~\bibnamefont {Ott}}, \ and\
		\bibinfo {author} {\bibfnamefont {J.~A.}\ \bibnamefont {Yorke}},\ }\bibfield
	{title} {\enquote {\bibinfo {title} {Crises, sudden changes in chaotic
				attractors, and transient chaos},}\ }\href@noop {} {\bibfield  {journal}
		{\bibinfo  {journal} {Physica D}\ }\textbf {\bibinfo {volume} {7}},\ \bibinfo
		{pages} {181--200} (\bibinfo {year} {1983})}\BibitemShut {NoStop}%
	\bibitem [{\citenamefont {Grebogi}\ \emph {et~al.}(1987)\citenamefont
		{Grebogi}, \citenamefont {Ott}, \citenamefont {Romeiras},\ and\ \citenamefont
		{Yorke}}]{grebogi1987critical}%
	\BibitemOpen
	\bibfield  {author} {\bibinfo {author} {\bibfnamefont {C.}~\bibnamefont
			{Grebogi}}, \bibinfo {author} {\bibfnamefont {E.}~\bibnamefont {Ott}},
		\bibinfo {author} {\bibfnamefont {F.}~\bibnamefont {Romeiras}}, \ and\
		\bibinfo {author} {\bibfnamefont {J.~A.}\ \bibnamefont {Yorke}},\ }\bibfield
	{title} {\enquote {\bibinfo {title} {Critical exponents for crisis-induced
				intermittency},}\ }\href {\doibase 10.1103/PhysRevA.36.5365} {\bibfield
		{journal} {\bibinfo  {journal} {Phys. Rev. A}\ }\textbf {\bibinfo {volume}
			{36}},\ \bibinfo {pages} {5365} (\bibinfo {year} {1987})}\BibitemShut
	{NoStop}%
	\bibitem [{\citenamefont {Pomeau}\ and\ \citenamefont
		{Manneville}(2017)}]{pomeau2017intermittent}%
	\BibitemOpen
	\bibfield  {author} {\bibinfo {author} {\bibfnamefont {Y.}~\bibnamefont
			{Pomeau}}\ and\ \bibinfo {author} {\bibfnamefont {P.}~\bibnamefont
			{Manneville}},\ }\bibfield  {title} {\enquote {\bibinfo {title} {Intermittent
				transition to turbulence in dissipative dynamical systems},}\ }in\ \href@noop
	{} {\emph {\bibinfo {booktitle} {Universality in Chaos}}}\ (\bibinfo
	{publisher} {Routledge},\ \bibinfo {year} {2017})\ pp.\ \bibinfo {pages}
	{327--335}\BibitemShut {NoStop}%
	\bibitem [{\citenamefont {Zamora-Munt}\ \emph {et~al.}(2013)\citenamefont
		{Zamora-Munt}, \citenamefont {Garbin}, \citenamefont {Barland}, \citenamefont
		{Giudici}, \citenamefont {Leite}, \citenamefont {Masoller},\ and\
		\citenamefont {Tredicce}}]{zamora2013rogue}%
	\BibitemOpen
	\bibfield  {author} {\bibinfo {author} {\bibfnamefont {J.}~\bibnamefont
			{Zamora-Munt}}, \bibinfo {author} {\bibfnamefont {B.}~\bibnamefont {Garbin}},
		\bibinfo {author} {\bibfnamefont {S.}~\bibnamefont {Barland}}, \bibinfo
		{author} {\bibfnamefont {M.}~\bibnamefont {Giudici}}, \bibinfo {author}
		{\bibfnamefont {J.~R.~R.}\ \bibnamefont {Leite}}, \bibinfo {author}
		{\bibfnamefont {C.}~\bibnamefont {Masoller}}, \ and\ \bibinfo {author}
		{\bibfnamefont {J.~R.}\ \bibnamefont {Tredicce}},\ }\bibfield  {title}
	{\enquote {\bibinfo {title} {Rogue waves in optically injected lasers:
				Origin, predictability, and suppression},}\ }\href@noop {} {\bibfield
		{journal} {\bibinfo  {journal} {Phys. Rev. A}\ }\textbf {\bibinfo {volume}
			{87}},\ \bibinfo {pages} {035802} (\bibinfo {year} {2013})}\BibitemShut
	{NoStop}%
	\bibitem [{\citenamefont {Kingston}\ \emph {et~al.}(2017)\citenamefont
		{Kingston}, \citenamefont {Thamilmaran}, \citenamefont {Pal}, \citenamefont
		{Feudel},\ and\ \citenamefont {Dana}}]{kingston2017extreme}%
	\BibitemOpen
	\bibfield  {author} {\bibinfo {author} {\bibfnamefont {S.~L.}\ \bibnamefont
			{Kingston}}, \bibinfo {author} {\bibfnamefont {K.}~\bibnamefont
			{Thamilmaran}}, \bibinfo {author} {\bibfnamefont {P.}~\bibnamefont {Pal}},
		\bibinfo {author} {\bibfnamefont {U.}~\bibnamefont {Feudel}}, \ and\ \bibinfo
		{author} {\bibfnamefont {S.~K.}\ \bibnamefont {Dana}},\ }\bibfield  {title}
	{\enquote {\bibinfo {title} {Extreme events in the forced li{\'e}nard
				system},}\ }\href@noop {} {\bibfield  {journal} {\bibinfo  {journal} {Phys.
				Rev. E}\ }\textbf {\bibinfo {volume} {96}},\ \bibinfo {pages} {052204}
		(\bibinfo {year} {2017})}\BibitemShut {NoStop}%
	\bibitem [{\citenamefont {Mishra}\ \emph {et~al.}(2020)\citenamefont {Mishra},
		\citenamefont {Leo~Kingston}, \citenamefont {Hens}, \citenamefont
		{Kapitaniak}, \citenamefont {Feudel},\ and\ \citenamefont
		{Dana}}]{mishra2020routes}%
	\BibitemOpen
	\bibfield  {author} {\bibinfo {author} {\bibfnamefont {A.}~\bibnamefont
			{Mishra}}, \bibinfo {author} {\bibfnamefont {S.}~\bibnamefont
			{Leo~Kingston}}, \bibinfo {author} {\bibfnamefont {C.}~\bibnamefont {Hens}},
		\bibinfo {author} {\bibfnamefont {T.}~\bibnamefont {Kapitaniak}}, \bibinfo
		{author} {\bibfnamefont {U.}~\bibnamefont {Feudel}}, \ and\ \bibinfo {author}
		{\bibfnamefont {S.~K.}\ \bibnamefont {Dana}},\ }\bibfield  {title} {\enquote
		{\bibinfo {title} {Routes to extreme events in dynamical systems: Dynamical
				and statistical characteristics},}\ }\href@noop {} {\bibfield  {journal}
		{\bibinfo  {journal} {Chaos}\ }\textbf {\bibinfo {volume} {30}},\ \bibinfo
		{pages} {063114} (\bibinfo {year} {2020})}\BibitemShut {NoStop}%
	\bibitem [{\citenamefont {Chowdhury}\ \emph {et~al.}(2022)\citenamefont
		{Chowdhury}, \citenamefont {Ray}, \citenamefont {Dana},\ and\ \citenamefont
		{Ghosh}}]{chowdhury2022extreme}%
	\BibitemOpen
	\bibfield  {author} {\bibinfo {author} {\bibfnamefont {S.~N.}\ \bibnamefont
			{Chowdhury}}, \bibinfo {author} {\bibfnamefont {A.}~\bibnamefont {Ray}},
		\bibinfo {author} {\bibfnamefont {S.~K.}\ \bibnamefont {Dana}}, \ and\
		\bibinfo {author} {\bibfnamefont {D.}~\bibnamefont {Ghosh}},\ }\bibfield
	{title} {\enquote {\bibinfo {title} {Extreme events in dynamical systems and
				random walkers: A review},}\ }\href@noop {} {\bibfield  {journal} {\bibinfo
			{journal} {Phys. Rep}\ }\textbf {\bibinfo {volume} {966}},\ \bibinfo {pages}
		{1--52} (\bibinfo {year} {2022})}\BibitemShut {NoStop}%
	\bibitem [{\citenamefont {Kaviya}\ \emph {et~al.}(2023)\citenamefont {Kaviya},
		\citenamefont {Gopal}, \citenamefont {Suresh},\ and\ \citenamefont
		{Chandrasekar}}]{kaviya2023route}%
	\BibitemOpen
	\bibfield  {author} {\bibinfo {author} {\bibfnamefont {B.}~\bibnamefont
			{Kaviya}}, \bibinfo {author} {\bibfnamefont {R.}~\bibnamefont {Gopal}},
		\bibinfo {author} {\bibfnamefont {R.}~\bibnamefont {Suresh}}, \ and\ \bibinfo
		{author} {\bibfnamefont {V.~K.}\ \bibnamefont {Chandrasekar}},\ }\bibfield
	{title} {\enquote {\bibinfo {title} {Route to extreme events in a
				parametrically driven position-dependent nonlinear oscillator},}\ }\href@noop
	{} {\bibfield  {journal} {\bibinfo  {journal} {Eur. Phys. J. Plus}\ }\textbf
		{\bibinfo {volume} {138}},\ \bibinfo {pages} {36} (\bibinfo {year}
		{2023})}\BibitemShut {NoStop}%
	\bibitem [{\citenamefont {Saha}\ and\ \citenamefont
		{Feudel}(2017)}]{saha2017extreme}%
	\BibitemOpen
	\bibfield  {author} {\bibinfo {author} {\bibfnamefont {A.}~\bibnamefont
			{Saha}}\ and\ \bibinfo {author} {\bibfnamefont {U.}~\bibnamefont {Feudel}},\
	}\bibfield  {title} {\enquote {\bibinfo {title} {Extreme events in
				fitzhugh-nagumo oscillators coupled with two time delays},}\ }\href@noop {}
	{\bibfield  {journal} {\bibinfo  {journal} {Phys. Rev. E}\ }\textbf {\bibinfo
			{volume} {95}},\ \bibinfo {pages} {062219} (\bibinfo {year}
		{2017})}\BibitemShut {NoStop}%
	\bibitem [{\citenamefont {Saha}\ and\ \citenamefont
		{Feudel}(2018)}]{saha2018riddled}%
	\BibitemOpen
	\bibfield  {author} {\bibinfo {author} {\bibfnamefont {A.}~\bibnamefont
			{Saha}}\ and\ \bibinfo {author} {\bibfnamefont {U.}~\bibnamefont {Feudel}},\
	}\bibfield  {title} {\enquote {\bibinfo {title} {Riddled basins of attraction
				in systems exhibiting extreme events},}\ }\href@noop {} {\bibfield  {journal}
		{\bibinfo  {journal} {Chaos}\ }\textbf {\bibinfo {volume} {28}} (\bibinfo
		{year} {2018})}\BibitemShut {NoStop}%
	\bibitem [{\citenamefont {Vijay}, \citenamefont {Thamilmaran},\ and\
		\citenamefont {Ahamed}(2023)}]{vijay2023superextreme}%
	\BibitemOpen
	\bibfield  {author} {\bibinfo {author} {\bibfnamefont {S.~D.}\ \bibnamefont
			{Vijay}}, \bibinfo {author} {\bibfnamefont {K.}~\bibnamefont {Thamilmaran}},
		\ and\ \bibinfo {author} {\bibfnamefont {A.~I.}\ \bibnamefont {Ahamed}},\
	}\bibfield  {title} {\enquote {\bibinfo {title} {Superextreme spiking
				oscillations and multistability in a memristor-based hindmarsh--rose neuron
				model},}\ }\href@noop {} {\bibfield  {journal} {\bibinfo  {journal}
			{Nonlinear Dyn}\ }\textbf {\bibinfo {volume} {111}},\ \bibinfo {pages}
		{789--799} (\bibinfo {year} {2023})}\BibitemShut {NoStop}%
	\bibitem [{\citenamefont {Thangavel}, \citenamefont {Srinivasan},\ and\
		\citenamefont {Kathamuthu}(2021)}]{thangavel2021extreme}%
	\BibitemOpen
	\bibfield  {author} {\bibinfo {author} {\bibfnamefont {B.}~\bibnamefont
			{Thangavel}}, \bibinfo {author} {\bibfnamefont {S.}~\bibnamefont
			{Srinivasan}}, \ and\ \bibinfo {author} {\bibfnamefont {T.}~\bibnamefont
			{Kathamuthu}},\ }\bibfield  {title} {\enquote {\bibinfo {title} {Extreme
				events in a forced bvp oscillator: Experimental and numerical studies},}\
	}\href@noop {} {\bibfield  {journal} {\bibinfo  {journal} {Chaos, Solitons \&
				Fractals}\ }\textbf {\bibinfo {volume} {153}},\ \bibinfo {pages} {111569}
		(\bibinfo {year} {2021})}\BibitemShut {NoStop}%
	\bibitem [{\citenamefont {Ray}\ \emph {et~al.}(2020)\citenamefont {Ray},
		\citenamefont {Mishra}, \citenamefont {Ghosh}, \citenamefont {Kapitaniak},
		\citenamefont {Dana},\ and\ \citenamefont {Hens}}]{ray2020extreme}%
	\BibitemOpen
	\bibfield  {author} {\bibinfo {author} {\bibfnamefont {A.}~\bibnamefont
			{Ray}}, \bibinfo {author} {\bibfnamefont {A.}~\bibnamefont {Mishra}},
		\bibinfo {author} {\bibfnamefont {D.}~\bibnamefont {Ghosh}}, \bibinfo
		{author} {\bibfnamefont {T.}~\bibnamefont {Kapitaniak}}, \bibinfo {author}
		{\bibfnamefont {S.~K.}\ \bibnamefont {Dana}}, \ and\ \bibinfo {author}
		{\bibfnamefont {C.}~\bibnamefont {Hens}},\ }\bibfield  {title} {\enquote
		{\bibinfo {title} {Extreme events in a network of heterogeneous josephson
				junctions},}\ }\href@noop {} {\bibfield  {journal} {\bibinfo  {journal}
			{Phys. Rev. E}\ }\textbf {\bibinfo {volume} {101}},\ \bibinfo {pages}
		{032209} (\bibinfo {year} {2020})}\BibitemShut {NoStop}%
	\bibitem [{\citenamefont {Fotopoulos}\ \emph {et~al.}(2020)\citenamefont
		{Fotopoulos}, \citenamefont {Frantzeskakis}, \citenamefont {Karachalios},
		\citenamefont {Kevrekidis}, \citenamefont {Koukouloyannis},\ and\
		\citenamefont {Vetas}}]{fotopoulos2020extreme}%
	\BibitemOpen
	\bibfield  {author} {\bibinfo {author} {\bibfnamefont {G.}~\bibnamefont
			{Fotopoulos}}, \bibinfo {author} {\bibfnamefont {D.~J.}\ \bibnamefont
			{Frantzeskakis}}, \bibinfo {author} {\bibfnamefont {N.~I.}\ \bibnamefont
			{Karachalios}}, \bibinfo {author} {\bibfnamefont {P.~G.}\ \bibnamefont
			{Kevrekidis}}, \bibinfo {author} {\bibfnamefont {V.}~\bibnamefont
			{Koukouloyannis}}, \ and\ \bibinfo {author} {\bibfnamefont {K.}~\bibnamefont
			{Vetas}},\ }\bibfield  {title} {\enquote {\bibinfo {title} {Extreme wave
				events for a nonlinear schr{\"o}dinger equation with linear damping and
				gaussian driving},}\ }\href@noop {} {\bibfield  {journal} {\bibinfo
			{journal} {Commun. Nonlinear Sci. Numer. Simul}\ }\textbf {\bibinfo {volume}
			{82}},\ \bibinfo {pages} {105058} (\bibinfo {year} {2020})}\BibitemShut
	{NoStop}%
	\bibitem [{\citenamefont {Kumarasamy}\ and\ \citenamefont
		{Pisarchik}(2018)}]{kumarasamy2018extreme}%
	\BibitemOpen
	\bibfield  {author} {\bibinfo {author} {\bibfnamefont {S.}~\bibnamefont
			{Kumarasamy}}\ and\ \bibinfo {author} {\bibfnamefont {A.~N.}\ \bibnamefont
			{Pisarchik}},\ }\bibfield  {title} {\enquote {\bibinfo {title} {Extreme
				events in systems with discontinuous boundaries},}\ }\href@noop {} {\bibfield
		{journal} {\bibinfo  {journal} {Phys. Rev. E}\ }\textbf {\bibinfo {volume}
			{98}},\ \bibinfo {pages} {032203} (\bibinfo {year} {2018})}\BibitemShut
	{NoStop}%
	\bibitem [{\citenamefont {Pal}\ \emph {et~al.}(2023)\citenamefont {Pal},
		\citenamefont {Ray}, \citenamefont {Nag~Chowdhury},\ and\ \citenamefont
		{Ghosh}}]{pal2023extreme}%
	\BibitemOpen
	\bibfield  {author} {\bibinfo {author} {\bibfnamefont {T.~K.}\ \bibnamefont
			{Pal}}, \bibinfo {author} {\bibfnamefont {A.}~\bibnamefont {Ray}}, \bibinfo
		{author} {\bibfnamefont {S.}~\bibnamefont {Nag~Chowdhury}}, \ and\ \bibinfo
		{author} {\bibfnamefont {D.}~\bibnamefont {Ghosh}},\ }\bibfield  {title}
	{\enquote {\bibinfo {title} {Extreme rotational events in a forced-damped
				nonlinear pendulum},}\ }\href@noop {} {\bibfield  {journal} {\bibinfo
			{journal} {Chaos}\ }\textbf {\bibinfo {volume} {33}},\ \bibinfo {pages}
		{063134} (\bibinfo {year} {2023})}\BibitemShut {NoStop}%
	\bibitem [{\citenamefont {Sudharsan}, \citenamefont {Venkatesan},\ and\
		\citenamefont {Senthilvelan}(2021)}]{sudharsan2021constant}%
	\BibitemOpen
	\bibfield  {author} {\bibinfo {author} {\bibfnamefont {S.}~\bibnamefont
			{Sudharsan}}, \bibinfo {author} {\bibfnamefont {A.}~\bibnamefont
			{Venkatesan}}, \ and\ \bibinfo {author} {\bibfnamefont {M.}~\bibnamefont
			{Senthilvelan}},\ }\bibfield  {title} {\enquote {\bibinfo {title} {Constant
				bias and weak second periodic forcing: tools to mitigate extreme events},}\
	}\href@noop {} {\bibfield  {journal} {\bibinfo  {journal} {Eur. Phys. J.
				Plus}\ }\textbf {\bibinfo {volume} {136}},\ \bibinfo {pages} {817} (\bibinfo
		{year} {2021})}\BibitemShut {NoStop}%
	\bibitem [{\citenamefont {Bonatto}\ and\ \citenamefont
		{Endler}(2017)}]{bonatto2017extreme}%
	\BibitemOpen
	\bibfield  {author} {\bibinfo {author} {\bibfnamefont {C.}~\bibnamefont
			{Bonatto}}\ and\ \bibinfo {author} {\bibfnamefont {A.}~\bibnamefont
			{Endler}},\ }\bibfield  {title} {\enquote {\bibinfo {title} {Extreme and
				superextreme events in a loss-modulated co 2 laser: Nonlinear resonance route
				and precursors},}\ }\href@noop {} {\bibfield  {journal} {\bibinfo  {journal}
			{Phys. Rev. E}\ }\textbf {\bibinfo {volume} {96}},\ \bibinfo {pages} {012216}
		(\bibinfo {year} {2017})}\BibitemShut {NoStop}%
	\bibitem [{\citenamefont {Sudharsan}\ \emph {et~al.}(2021)\citenamefont
		{Sudharsan}, \citenamefont {Venkatesan}, \citenamefont {Muruganandam},\ and\
		\citenamefont {Senthilvelan}}]{sudharsan2021emergence}%
	\BibitemOpen
	\bibfield  {author} {\bibinfo {author} {\bibfnamefont {S.}~\bibnamefont
			{Sudharsan}}, \bibinfo {author} {\bibfnamefont {A.}~\bibnamefont
			{Venkatesan}}, \bibinfo {author} {\bibfnamefont {P.}~\bibnamefont
			{Muruganandam}}, \ and\ \bibinfo {author} {\bibfnamefont {M.}~\bibnamefont
			{Senthilvelan}},\ }\bibfield  {title} {\enquote {\bibinfo {title} {Emergence
				and mitigation of extreme events in a parametrically driven system with
				velocity-dependent potential},}\ }\href@noop {} {\bibfield  {journal}
		{\bibinfo  {journal} {Eur. Phys. J. Plus}\ }\textbf {\bibinfo {volume}
			{136}},\ \bibinfo {pages} {129} (\bibinfo {year} {2021})}\BibitemShut
	{NoStop}%
	\bibitem [{\citenamefont {Sudharsan}\ \emph {et~al.}(2022)\citenamefont
		{Sudharsan}, \citenamefont {Venkatesan}, \citenamefont {Muruganandam},\ and\
		\citenamefont {Senthilvelan}}]{sudharsan2022suppression}%
	\BibitemOpen
	\bibfield  {author} {\bibinfo {author} {\bibfnamefont {S.}~\bibnamefont
			{Sudharsan}}, \bibinfo {author} {\bibfnamefont {A.}~\bibnamefont
			{Venkatesan}}, \bibinfo {author} {\bibfnamefont {P.}~\bibnamefont
			{Muruganandam}}, \ and\ \bibinfo {author} {\bibfnamefont {M.}~\bibnamefont
			{Senthilvelan}},\ }\bibfield  {title} {\enquote {\bibinfo {title}
			{Suppression of extreme events and chaos in a velocity-dependent potential
				system with time-delay feedback},}\ }\href@noop {} {\bibfield  {journal}
		{\bibinfo  {journal} {Chaos, Solitons \& Fractals}\ }\textbf {\bibinfo
			{volume} {161}},\ \bibinfo {pages} {112321} (\bibinfo {year}
		{2022})}\BibitemShut {NoStop}%
	\bibitem [{\citenamefont {Roy}\ \emph {et~al.}(2022)\citenamefont {Roy},
		\citenamefont {Mandal}, \citenamefont {Hens}, \citenamefont {Prasad},
		\citenamefont {Kuznetsov},\ and\ \citenamefont
		{Dev~Shrimali}}]{roy2022model}%
	\BibitemOpen
	\bibfield  {author} {\bibinfo {author} {\bibfnamefont {M.}~\bibnamefont
			{Roy}}, \bibinfo {author} {\bibfnamefont {S.}~\bibnamefont {Mandal}},
		\bibinfo {author} {\bibfnamefont {C.}~\bibnamefont {Hens}}, \bibinfo {author}
		{\bibfnamefont {A.}~\bibnamefont {Prasad}}, \bibinfo {author} {\bibfnamefont
			{N.}~\bibnamefont {Kuznetsov}}, \ and\ \bibinfo {author} {\bibfnamefont
			{M.}~\bibnamefont {Dev~Shrimali}},\ }\bibfield  {title} {\enquote {\bibinfo
			{title} {Model-free prediction of multistability using echo state network},}\
	}\href@noop {} {\bibfield  {journal} {\bibinfo  {journal} {Chaos}\ }\textbf
		{\bibinfo {volume} {32}} (\bibinfo {year} {2022})}\BibitemShut {NoStop}%
	\bibitem [{\citenamefont {Shashangan}\ \emph {et~al.}(2024)\citenamefont
		{Shashangan}, \citenamefont {Sudharsan}, \citenamefont {Venkatesan},\ and\
		\citenamefont {Senthilvelan}}]{shashangan2024mitigation}%
	\BibitemOpen
	\bibfield  {author} {\bibinfo {author} {\bibfnamefont {R.}~\bibnamefont
			{Shashangan}}, \bibinfo {author} {\bibfnamefont {S.}~\bibnamefont
			{Sudharsan}}, \bibinfo {author} {\bibfnamefont {A.}~\bibnamefont
			{Venkatesan}}, \ and\ \bibinfo {author} {\bibfnamefont {M.}~\bibnamefont
			{Senthilvelan}},\ }\bibfield  {title} {\enquote {\bibinfo {title} {Mitigation
				of extreme events in an excitable system},}\ }\href@noop {} {\bibfield
		{journal} {\bibinfo  {journal} {Eur. Phys. J. Plus}\ }\textbf {\bibinfo
			{volume} {139}},\ \bibinfo {pages} {203} (\bibinfo {year}
		{2024})}\BibitemShut {NoStop}%
	\bibitem [{\citenamefont {Bastard}(1990)}]{bastard1990wave}%
	\BibitemOpen
	\bibfield  {author} {\bibinfo {author} {\bibfnamefont {G.}~\bibnamefont
			{Bastard}},\ }\href@noop {} {\emph {\bibinfo {title} {Wave mechanics applied
				to semiconductor heterostructures}}}\ (\bibinfo  {publisher} {New York, NY
		(USA); John Wiley and Sons Inc.},\ \bibinfo {year} {1990})\BibitemShut
	{NoStop}%
	\bibitem [{\citenamefont {G{\"o}n{\"u}l}\ \emph {et~al.}(2002)\citenamefont
		{G{\"o}n{\"u}l}, \citenamefont {{\"O}zer}, \citenamefont {G{\"o}n{\"u}L},\
		and\ \citenamefont {{\"U}zg{\"u}n}}]{gonul2002exact}%
	\BibitemOpen
	\bibfield  {author} {\bibinfo {author} {\bibfnamefont {B.}~\bibnamefont
			{G{\"o}n{\"u}l}}, \bibinfo {author} {\bibfnamefont {O.}~\bibnamefont
			{{\"O}zer}}, \bibinfo {author} {\bibfnamefont {B.}~\bibnamefont
			{G{\"o}n{\"u}L}}, \ and\ \bibinfo {author} {\bibfnamefont {F.}~\bibnamefont
			{{\"U}zg{\"u}n}},\ }\bibfield  {title} {\enquote {\bibinfo {title} {Exact
				solutions of effective-mass schr{\"o}dinger equations},}\ }\href@noop {}
	{\bibfield  {journal} {\bibinfo  {journal} {Modern Physics Letters A}\
		}\textbf {\bibinfo {volume} {17}},\ \bibinfo {pages} {2453--2465} (\bibinfo
		{year} {2002})}\BibitemShut {NoStop}%
	\bibitem [{\citenamefont {Jordan}\ and\ \citenamefont
		{Mitchell}(2015)}]{jordan2015machine}%
	\BibitemOpen
	\bibfield  {author} {\bibinfo {author} {\bibfnamefont {M.~I.}\ \bibnamefont
			{Jordan}}\ and\ \bibinfo {author} {\bibfnamefont {T.~M.}\ \bibnamefont
			{Mitchell}},\ }\bibfield  {title} {\enquote {\bibinfo {title} {Machine
				learning: Trends, perspectives, and prospects},}\ }\href@noop {} {\bibfield
		{journal} {\bibinfo  {journal} {Science}\ }\textbf {\bibinfo {volume}
			{349}},\ \bibinfo {pages} {255--260} (\bibinfo {year} {2015})}\BibitemShut
	{NoStop}%
	\bibitem [{\citenamefont {Mohri}, \citenamefont {Rostamizadeh},\ and\
		\citenamefont {Talwalkar}(2018)}]{mohri2018foundations}%
	\BibitemOpen
	\bibfield  {author} {\bibinfo {author} {\bibfnamefont {M.}~\bibnamefont
			{Mohri}}, \bibinfo {author} {\bibfnamefont {A.}~\bibnamefont {Rostamizadeh}},
		\ and\ \bibinfo {author} {\bibfnamefont {A.}~\bibnamefont {Talwalkar}},\
	}\href@noop {} {\emph {\bibinfo {title} {Foundations of machine learning}}}\
	(\bibinfo  {publisher} {MIT press},\ \bibinfo {year} {2018})\BibitemShut
	{NoStop}%
	\bibitem [{\citenamefont {Carleo}\ \emph {et~al.}(2019)\citenamefont {Carleo},
		\citenamefont {Cirac}, \citenamefont {Cranmer}, \citenamefont {Daudet},
		\citenamefont {Schuld}, \citenamefont {Tishby}, \citenamefont
		{Vogt-Maranto},\ and\ \citenamefont {Zdeborov{\'a}}}]{carleo2019machine}%
	\BibitemOpen
	\bibfield  {author} {\bibinfo {author} {\bibfnamefont {G.}~\bibnamefont
			{Carleo}}, \bibinfo {author} {\bibfnamefont {I.}~\bibnamefont {Cirac}},
		\bibinfo {author} {\bibfnamefont {K.}~\bibnamefont {Cranmer}}, \bibinfo
		{author} {\bibfnamefont {L.}~\bibnamefont {Daudet}}, \bibinfo {author}
		{\bibfnamefont {M.}~\bibnamefont {Schuld}}, \bibinfo {author} {\bibfnamefont
			{N.}~\bibnamefont {Tishby}}, \bibinfo {author} {\bibfnamefont
			{L.}~\bibnamefont {Vogt-Maranto}}, \ and\ \bibinfo {author} {\bibfnamefont
			{L.}~\bibnamefont {Zdeborov{\'a}}},\ }\bibfield  {title} {\enquote {\bibinfo
			{title} {Machine learning and the physical sciences},}\ }\href@noop {}
	{\bibfield  {journal} {\bibinfo  {journal} {Rev. Mod. Phys.}\ }\textbf
		{\bibinfo {volume} {91}},\ \bibinfo {pages} {045002} (\bibinfo {year}
		{2019})}\BibitemShut {NoStop}%
	\bibitem [{\citenamefont {Alpaydin}(2020)}]{alpaydin2020introduction}%
	\BibitemOpen
	\bibfield  {author} {\bibinfo {author} {\bibfnamefont {E.}~\bibnamefont
			{Alpaydin}},\ }\href@noop {} {\emph {\bibinfo {title} {Introduction to
				machine learning}}}\ (\bibinfo  {publisher} {MIT press},\ \bibinfo {year}
	{2020})\BibitemShut {NoStop}%
	\bibitem [{\citenamefont {Alpaydin}(2021)}]{alpaydin2021machine}%
	\BibitemOpen
	\bibfield  {author} {\bibinfo {author} {\bibfnamefont {E.}~\bibnamefont
			{Alpaydin}},\ }\href@noop {} {\emph {\bibinfo {title} {Machine learning}}}\
	(\bibinfo  {publisher} {MIT press},\ \bibinfo {year} {2021})\BibitemShut
	{NoStop}%
	\bibitem [{\citenamefont {Ljung}(2010)}]{ljung2010approaches}%
	\BibitemOpen
	\bibfield  {author} {\bibinfo {author} {\bibfnamefont {L.}~\bibnamefont
			{Ljung}},\ }\bibfield  {title} {\enquote {\bibinfo {title} {Approaches to
				identification of nonlinear systems},}\ }in\ \href@noop {} {\emph {\bibinfo
			{booktitle} {Proceedings of the 29th Chinese Control Conference}}}\ (\bibinfo
	{organization} {IEEE},\ \bibinfo {year} {2010})\ pp.\ \bibinfo {pages}
	{1--5}\BibitemShut {NoStop}%
	\bibitem [{\citenamefont {Brunton}, \citenamefont {Proctor},\ and\
		\citenamefont {Kutz}(2016)}]{brunton2016discovering}%
	\BibitemOpen
	\bibfield  {author} {\bibinfo {author} {\bibfnamefont {S.~L.}\ \bibnamefont
			{Brunton}}, \bibinfo {author} {\bibfnamefont {J.~L.}\ \bibnamefont
			{Proctor}}, \ and\ \bibinfo {author} {\bibfnamefont {J.~N.}\ \bibnamefont
			{Kutz}},\ }\bibfield  {title} {\enquote {\bibinfo {title} {Discovering
				governing equations from data by sparse identification of nonlinear dynamical
				systems},}\ }\href@noop {} {\bibfield  {journal} {\bibinfo  {journal}
			{Proceedings of the National Academy of Sciences}\ }\textbf {\bibinfo
			{volume} {113}},\ \bibinfo {pages} {3932--3937} (\bibinfo {year}
		{2016})}\BibitemShut {NoStop}%
	\bibitem [{\citenamefont {Tang}\ \emph {et~al.}(2020)\citenamefont {Tang},
		\citenamefont {Kurths}, \citenamefont {Lin}, \citenamefont {Ott},\ and\
		\citenamefont {Kocarev}}]{tang2020introduction}%
	\BibitemOpen
	\bibfield  {author} {\bibinfo {author} {\bibfnamefont {Y.}~\bibnamefont
			{Tang}}, \bibinfo {author} {\bibfnamefont {J.}~\bibnamefont {Kurths}},
		\bibinfo {author} {\bibfnamefont {W.}~\bibnamefont {Lin}}, \bibinfo {author}
		{\bibfnamefont {E.}~\bibnamefont {Ott}}, \ and\ \bibinfo {author}
		{\bibfnamefont {L.}~\bibnamefont {Kocarev}},\ }\bibfield  {title} {\enquote
		{\bibinfo {title} {Introduction to focus issue: When machine learning meets
				complex systems: Networks, chaos, and nonlinear dynamics},}\ }\href@noop {}
	{\bibfield  {journal} {\bibinfo  {journal} {Chaos}\ }\textbf {\bibinfo
			{volume} {30}} (\bibinfo {year} {2020})}\BibitemShut {NoStop}%
	\bibitem [{\citenamefont {Pourmohammad~Azizi}, \citenamefont {Neisy},\ and\
		\citenamefont {Ahmad~Waloo}(2023)}]{pourmohammad2023dynamical}%
	\BibitemOpen
	\bibfield  {author} {\bibinfo {author} {\bibfnamefont {S.}~\bibnamefont
			{Pourmohammad~Azizi}}, \bibinfo {author} {\bibfnamefont {A.}~\bibnamefont
			{Neisy}}, \ and\ \bibinfo {author} {\bibfnamefont {S.}~\bibnamefont
			{Ahmad~Waloo}},\ }\bibfield  {title} {\enquote {\bibinfo {title} {A dynamical
				systems approach to machine learning},}\ }\href@noop {} {\bibfield  {journal}
		{\bibinfo  {journal} {Int. J. Comput. Methods}\ }\textbf {\bibinfo {volume}
			{20}},\ \bibinfo {pages} {2350007} (\bibinfo {year} {2023})}\BibitemShut
	{NoStop}%
	\bibitem [{\citenamefont {Pan}\ and\ \citenamefont
		{Wang}(2011)}]{pan2011model}%
	\BibitemOpen
	\bibfield  {author} {\bibinfo {author} {\bibfnamefont {Y.}~\bibnamefont
			{Pan}}\ and\ \bibinfo {author} {\bibfnamefont {J.}~\bibnamefont {Wang}},\
	}\bibfield  {title} {\enquote {\bibinfo {title} {Model predictive control of
				unknown nonlinear dynamical systems based on recurrent neural networks},}\
	}\href@noop {} {\bibfield  {journal} {\bibinfo  {journal} {IEEE Trans. Ind.
				Electron}\ }\textbf {\bibinfo {volume} {59}},\ \bibinfo {pages} {3089--3101}
		(\bibinfo {year} {2011})}\BibitemShut {NoStop}%
	\bibitem [{\citenamefont {Liu}\ \emph {et~al.}(2023)\citenamefont {Liu},
		\citenamefont {Peng}, \citenamefont {Long}, \citenamefont {Wang},
		\citenamefont {Yang}, \citenamefont {P{\'e}rez-Jim{\'e}nez},\ and\
		\citenamefont {Orellana-Mart{\'\i}n}}]{liu2023nonlinear}%
	\BibitemOpen
	\bibfield  {author} {\bibinfo {author} {\bibfnamefont {Q.}~\bibnamefont
			{Liu}}, \bibinfo {author} {\bibfnamefont {H.}~\bibnamefont {Peng}}, \bibinfo
		{author} {\bibfnamefont {L.}~\bibnamefont {Long}}, \bibinfo {author}
		{\bibfnamefont {J.}~\bibnamefont {Wang}}, \bibinfo {author} {\bibfnamefont
			{Q.}~\bibnamefont {Yang}}, \bibinfo {author} {\bibfnamefont {M.~J.}\
			\bibnamefont {P{\'e}rez-Jim{\'e}nez}}, \ and\ \bibinfo {author}
		{\bibfnamefont {D.}~\bibnamefont {Orellana-Mart{\'\i}n}},\ }\bibfield
	{title} {\enquote {\bibinfo {title} {Nonlinear spiking neural systems with
				autapses for predicting chaotic time series},}\ }\href@noop {} {\bibfield
		{journal} {\bibinfo  {journal} {IEEE Trans. Cybern.}\ } (\bibinfo {year}
		{2023})}\BibitemShut {NoStop}%
	\bibitem [{\citenamefont {Meiyazhagan}, \citenamefont {Sudharsan},\ and\
		\citenamefont {Senthilvelan}(2021)}]{meiyazhagan2021model}%
	\BibitemOpen
	\bibfield  {author} {\bibinfo {author} {\bibfnamefont {J.}~\bibnamefont
			{Meiyazhagan}}, \bibinfo {author} {\bibfnamefont {S.}~\bibnamefont
			{Sudharsan}}, \ and\ \bibinfo {author} {\bibfnamefont {M.}~\bibnamefont
			{Senthilvelan}},\ }\bibfield  {title} {\enquote {\bibinfo {title} {Model-free
				prediction of emergence of extreme events in a parametrically driven
				nonlinear dynamical system by deep learning},}\ }\href@noop {} {\bibfield
		{journal} {\bibinfo  {journal} {Eur. Phys. J. B}\ }\textbf {\bibinfo {volume}
			{94}},\ \bibinfo {pages} {156} (\bibinfo {year} {2021})}\BibitemShut
	{NoStop}%
	\bibitem [{\citenamefont {Chen}\ \emph {et~al.}(2022)\citenamefont {Chen},
		\citenamefont {Jin}, \citenamefont {Laima}, \citenamefont {Huang},\ and\
		\citenamefont {Li}}]{chen2022intelligent}%
	\BibitemOpen
	\bibfield  {author} {\bibinfo {author} {\bibfnamefont {R.}~\bibnamefont
			{Chen}}, \bibinfo {author} {\bibfnamefont {X.}~\bibnamefont {Jin}}, \bibinfo
		{author} {\bibfnamefont {S.}~\bibnamefont {Laima}}, \bibinfo {author}
		{\bibfnamefont {Y.}~\bibnamefont {Huang}}, \ and\ \bibinfo {author}
		{\bibfnamefont {H.}~\bibnamefont {Li}},\ }\bibfield  {title} {\enquote
		{\bibinfo {title} {Intelligent modeling of nonlinear dynamical systems by
				machine learning},}\ }\href@noop {} {\bibfield  {journal} {\bibinfo
			{journal} {Int. J. Non-Linear Mech}\ }\textbf {\bibinfo {volume} {142}},\
		\bibinfo {pages} {103984} (\bibinfo {year} {2022})}\BibitemShut {NoStop}%
	\bibitem [{\citenamefont {Meiyazhagan}\ \emph {et~al.}(2021)\citenamefont
		{Meiyazhagan}, \citenamefont {Sudharsan}, \citenamefont {Venkatesan},\ and\
		\citenamefont {Senthilvelan}}]{meiyazhagan2021prediction}%
	\BibitemOpen
	\bibfield  {author} {\bibinfo {author} {\bibfnamefont {J.}~\bibnamefont
			{Meiyazhagan}}, \bibinfo {author} {\bibfnamefont {S.}~\bibnamefont
			{Sudharsan}}, \bibinfo {author} {\bibfnamefont {A.}~\bibnamefont
			{Venkatesan}}, \ and\ \bibinfo {author} {\bibfnamefont {M.}~\bibnamefont
			{Senthilvelan}},\ }\bibfield  {title} {\enquote {\bibinfo {title} {Prediction
				of occurrence of extreme events using machine learning},}\ }\href@noop {}
	{\bibfield  {journal} {\bibinfo  {journal} {Eur. Phys. J. Plus}\ }\textbf
		{\bibinfo {volume} {137}},\ \bibinfo {pages} {16} (\bibinfo {year}
		{2021})}\BibitemShut {NoStop}%
	\bibitem [{\citenamefont {Jaeger}\ and\ \citenamefont
		{Haas}(2004)}]{jaeger2004harnessing}%
	\BibitemOpen
	\bibfield  {author} {\bibinfo {author} {\bibfnamefont {H.}~\bibnamefont
			{Jaeger}}\ and\ \bibinfo {author} {\bibfnamefont {H.}~\bibnamefont {Haas}},\
	}\bibfield  {title} {\enquote {\bibinfo {title} {Harnessing nonlinearity:
				Predicting chaotic systems and saving energy in wireless communication},}\
	}\href@noop {} {\bibfield  {journal} {\bibinfo  {journal} {Science}\ }\textbf
		{\bibinfo {volume} {304}},\ \bibinfo {pages} {78--80} (\bibinfo {year}
		{2004})}\BibitemShut {NoStop}%
	\bibitem [{\citenamefont {Pathak}\ and\ \citenamefont
		{Ott}(2021)}]{pathak2021reservoir}%
	\BibitemOpen
	\bibfield  {author} {\bibinfo {author} {\bibfnamefont {J.}~\bibnamefont
			{Pathak}}\ and\ \bibinfo {author} {\bibfnamefont {E.}~\bibnamefont {Ott}},\
	}\bibfield  {title} {\enquote {\bibinfo {title} {Reservoir computing for
				forecasting large spatiotemporal dynamical systems},}\ }\href@noop {}
	{\bibfield  {journal} {\bibinfo  {journal} {Reservoir Computing: Theory,
				Physical Implementations, and Applications}\ ,\ \bibinfo {pages} {117--138}}
		(\bibinfo {year} {2021})}\BibitemShut {NoStop}%
	\bibitem [{\citenamefont {Barmparis}\ \emph {et~al.}(2020)\citenamefont
		{Barmparis}, \citenamefont {Neofotistos}, \citenamefont {Mattheakis},
		\citenamefont {Hizanidis}, \citenamefont {Tsironis},\ and\ \citenamefont
		{Kaxiras}}]{BARMPARIS2020126300}%
	\BibitemOpen
	\bibfield  {author} {\bibinfo {author} {\bibfnamefont {G.}~\bibnamefont
			{Barmparis}}, \bibinfo {author} {\bibfnamefont {G.}~\bibnamefont
			{Neofotistos}}, \bibinfo {author} {\bibfnamefont {M.}~\bibnamefont
			{Mattheakis}}, \bibinfo {author} {\bibfnamefont {J.}~\bibnamefont
			{Hizanidis}}, \bibinfo {author} {\bibfnamefont {G.}~\bibnamefont {Tsironis}},
		\ and\ \bibinfo {author} {\bibfnamefont {E.}~\bibnamefont {Kaxiras}},\
	}\bibfield  {title} {\enquote {\bibinfo {title} {Robust prediction of complex
				spatiotemporal states through machine learning with sparse sensing},}\ }\href
	{\doibase https://doi.org/10.1016/j.physleta.2020.126300} {\bibfield
		{journal} {\bibinfo  {journal} {Phys. Lett. A}\ }\textbf {\bibinfo {volume}
			{384}},\ \bibinfo {pages} {126300} (\bibinfo {year} {2020})}\BibitemShut
	{NoStop}%
	\bibitem [{\citenamefont {Kushwaha}\ \emph {et~al.}(2021)\citenamefont
		{Kushwaha}, \citenamefont {Mendola}, \citenamefont {Ghosh}, \citenamefont
		{Kachhvah},\ and\ \citenamefont {Jalan}}]{Kushwaha_2021}%
	\BibitemOpen
	\bibfield  {author} {\bibinfo {author} {\bibfnamefont {N.}~\bibnamefont
			{Kushwaha}}, \bibinfo {author} {\bibfnamefont {N.~K.}\ \bibnamefont
			{Mendola}}, \bibinfo {author} {\bibfnamefont {S.}~\bibnamefont {Ghosh}},
		\bibinfo {author} {\bibfnamefont {A.~D.}\ \bibnamefont {Kachhvah}}, \ and\
		\bibinfo {author} {\bibfnamefont {S.}~\bibnamefont {Jalan}},\ }\bibfield
	{title} {\enquote {\bibinfo {title} {Machine learning assisted chimera and
				solitary states in networks},}\ }\href {\doibase 10.3389/fphy.2021.513969}
	{\bibfield  {journal} {\bibinfo  {journal} {Front. Phys.}\ }\textbf {\bibinfo
			{volume} {9}} (\bibinfo {year} {2021}),\
		10.3389/fphy.2021.513969}\BibitemShut {NoStop}%
	\bibitem [{\citenamefont {Ganaie}\ \emph {et~al.}(2020)\citenamefont {Ganaie},
		\citenamefont {Ghosh}, \citenamefont {Mendola}, \citenamefont {Tanveer},\
		and\ \citenamefont {Jalan}}]{ganaie2020identification}%
	\BibitemOpen
	\bibfield  {author} {\bibinfo {author} {\bibfnamefont {M.}~\bibnamefont
			{Ganaie}}, \bibinfo {author} {\bibfnamefont {S.}~\bibnamefont {Ghosh}},
		\bibinfo {author} {\bibfnamefont {N.}~\bibnamefont {Mendola}}, \bibinfo
		{author} {\bibfnamefont {M.}~\bibnamefont {Tanveer}}, \ and\ \bibinfo
		{author} {\bibfnamefont {S.}~\bibnamefont {Jalan}},\ }\bibfield  {title}
	{\enquote {\bibinfo {title} {Identification of chimera using machine
				learning},}\ }\href@noop {} {\bibfield  {journal} {\bibinfo  {journal}
			{Chaos}\ }\textbf {\bibinfo {volume} {30}} (\bibinfo {year}
		{2020})}\BibitemShut {NoStop}%
	\bibitem [{\citenamefont {Xiao}\ \emph {et~al.}(2021)\citenamefont {Xiao},
		\citenamefont {Kong}, \citenamefont {Sun},\ and\ \citenamefont
		{Lai}}]{xiao2021predicting}%
	\BibitemOpen
	\bibfield  {author} {\bibinfo {author} {\bibfnamefont {R.}~\bibnamefont
			{Xiao}}, \bibinfo {author} {\bibfnamefont {L.-W.}\ \bibnamefont {Kong}},
		\bibinfo {author} {\bibfnamefont {Z.-K.}\ \bibnamefont {Sun}}, \ and\
		\bibinfo {author} {\bibfnamefont {Y.-C.}\ \bibnamefont {Lai}},\ }\bibfield
	{title} {\enquote {\bibinfo {title} {Predicting amplitude death with machine
				learning},}\ }\href@noop {} {\bibfield  {journal} {\bibinfo  {journal} {Phys.
				Rev. E}\ }\textbf {\bibinfo {volume} {104}},\ \bibinfo {pages} {014205}
		(\bibinfo {year} {2021})}\BibitemShut {NoStop}%
	\bibitem [{\citenamefont {Ray}, \citenamefont {Chakraborty},\ and\
		\citenamefont {Ghosh}(2021)}]{ray2021optimized}%
	\BibitemOpen
	\bibfield  {author} {\bibinfo {author} {\bibfnamefont {A.}~\bibnamefont
			{Ray}}, \bibinfo {author} {\bibfnamefont {T.}~\bibnamefont {Chakraborty}}, \
		and\ \bibinfo {author} {\bibfnamefont {D.}~\bibnamefont {Ghosh}},\ }\bibfield
	{title} {\enquote {\bibinfo {title} {Optimized ensemble deep learning
				framework for scalable forecasting of dynamics containing extreme events},}\
	}\href@noop {} {\bibfield  {journal} {\bibinfo  {journal} {Chaos}\ }\textbf
		{\bibinfo {volume} {31}},\ \bibinfo {pages} {111105} (\bibinfo {year}
		{2021})}\BibitemShut {NoStop}%
	\bibitem [{\citenamefont {Pyragas}\ and\ \citenamefont
		{Pyragas}(2020)}]{PYRAGAS2020126591}%
	\BibitemOpen
	\bibfield  {author} {\bibinfo {author} {\bibfnamefont {V.}~\bibnamefont
			{Pyragas}}\ and\ \bibinfo {author} {\bibfnamefont {K.}~\bibnamefont
			{Pyragas}},\ }\bibfield  {title} {\enquote {\bibinfo {title} {Using reservoir
				computer to predict and prevent extreme events},}\ }\href {\doibase
		https://doi.org/10.1016/j.physleta.2020.126591} {\bibfield  {journal}
		{\bibinfo  {journal} {Phys. Lett. A}\ }\textbf {\bibinfo {volume} {384}},\
		\bibinfo {pages} {126591} (\bibinfo {year} {2020})}\BibitemShut {NoStop}%
	\bibitem [{\citenamefont {Durairaj}\ \emph {et~al.}(2023)\citenamefont
		{Durairaj}, \citenamefont {Sundararam}, \citenamefont {Kanagaraj},\ and\
		\citenamefont {Rajagopal}}]{durairaj2023prediction}%
	\BibitemOpen
	\bibfield  {author} {\bibinfo {author} {\bibfnamefont {P.}~\bibnamefont
			{Durairaj}}, \bibinfo {author} {\bibfnamefont {G.~K.}\ \bibnamefont
			{Sundararam}}, \bibinfo {author} {\bibfnamefont {S.}~\bibnamefont
			{Kanagaraj}}, \ and\ \bibinfo {author} {\bibfnamefont {K.}~\bibnamefont
			{Rajagopal}},\ }\bibfield  {title} {\enquote {\bibinfo {title} {Prediction of
				dragon king extreme events using machine learning approaches and its
				characterizations},}\ }\href@noop {} {\bibfield  {journal} {\bibinfo
			{journal} {Phys. Lett. A}\ }\textbf {\bibinfo {volume} {489}},\ \bibinfo
		{pages} {129158} (\bibinfo {year} {2023})}\BibitemShut {NoStop}%
	\bibitem [{\citenamefont {Jiang}\ \emph {et~al.}(2022)\citenamefont {Jiang},
		\citenamefont {Huang}, \citenamefont {Grebogi},\ and\ \citenamefont
		{Lai}}]{jiang2022predicting}%
	\BibitemOpen
	\bibfield  {author} {\bibinfo {author} {\bibfnamefont {J.}~\bibnamefont
			{Jiang}}, \bibinfo {author} {\bibfnamefont {Z.-G.}\ \bibnamefont {Huang}},
		\bibinfo {author} {\bibfnamefont {C.}~\bibnamefont {Grebogi}}, \ and\
		\bibinfo {author} {\bibfnamefont {Y.-C.}\ \bibnamefont {Lai}},\ }\bibfield
	{title} {\enquote {\bibinfo {title} {Predicting extreme events from data
				using deep machine learning: When and where},}\ }\href@noop {} {\bibfield
		{journal} {\bibinfo  {journal} {Phys. Rev. Res.}\ }\textbf {\bibinfo {volume}
			{4}},\ \bibinfo {pages} {023028} (\bibinfo {year} {2022})}\BibitemShut
	{NoStop}%
	\bibitem [{\citenamefont {Grebogi}, \citenamefont {Ott},\ and\ \citenamefont
		{Yorke}(1987)}]{grebogi1987unstable}%
	\BibitemOpen
	\bibfield  {author} {\bibinfo {author} {\bibfnamefont {C.}~\bibnamefont
			{Grebogi}}, \bibinfo {author} {\bibfnamefont {E.}~\bibnamefont {Ott}}, \ and\
		\bibinfo {author} {\bibfnamefont {J.~A.}\ \bibnamefont {Yorke}},\ }\bibfield
	{title} {\enquote {\bibinfo {title} {Unstable periodic orbits and the
				dimension of chaotic attractors},}\ }\href@noop {} {\bibfield  {journal}
		{\bibinfo  {journal} {Phys. Rev. A}\ }\textbf {\bibinfo {volume} {36}},\
		\bibinfo {pages} {3522} (\bibinfo {year} {1987})}\BibitemShut {NoStop}%
	\bibitem [{\citenamefont {Wolf}\ \emph {et~al.}(1985)\citenamefont {Wolf},
		\citenamefont {Swift}, \citenamefont {Swinney},\ and\ \citenamefont
		{Vastano}}]{wolf1985determining}%
	\BibitemOpen
	\bibfield  {author} {\bibinfo {author} {\bibfnamefont {A.}~\bibnamefont
			{Wolf}}, \bibinfo {author} {\bibfnamefont {J.~B.}\ \bibnamefont {Swift}},
		\bibinfo {author} {\bibfnamefont {H.~L.}\ \bibnamefont {Swinney}}, \ and\
		\bibinfo {author} {\bibfnamefont {J.~A.}\ \bibnamefont {Vastano}},\
	}\bibfield  {title} {\enquote {\bibinfo {title} {Determining lyapunov
				exponents from a time series},}\ }\href@noop {} {\bibfield  {journal}
		{\bibinfo  {journal} {Physica D}\ }\textbf {\bibinfo {volume} {16}},\
		\bibinfo {pages} {285--317} (\bibinfo {year} {1985})}\BibitemShut {NoStop}%
	\bibitem [{\citenamefont {Grebogi}, \citenamefont {Ott},\ and\ \citenamefont
		{Yorke}(1982)}]{grebogi1982chaotic}%
	\BibitemOpen
	\bibfield  {author} {\bibinfo {author} {\bibfnamefont {C.}~\bibnamefont
			{Grebogi}}, \bibinfo {author} {\bibfnamefont {E.}~\bibnamefont {Ott}}, \ and\
		\bibinfo {author} {\bibfnamefont {J.~A.}\ \bibnamefont {Yorke}},\ }\bibfield
	{title} {\enquote {\bibinfo {title} {Chaotic attractors in crisis},}\
	}\href@noop {} {\bibfield  {journal} {\bibinfo  {journal} {Phys. Rev. Lett}\
		}\textbf {\bibinfo {volume} {48}},\ \bibinfo {pages} {1507} (\bibinfo {year}
		{1982})}\BibitemShut {NoStop}%
	\bibitem [{\citenamefont {Chossat}\ and\ \citenamefont
		{Golubitsky}(1988)}]{chossat1988symmetry}%
	\BibitemOpen
	\bibfield  {author} {\bibinfo {author} {\bibfnamefont {P.}~\bibnamefont
			{Chossat}}\ and\ \bibinfo {author} {\bibfnamefont {M.}~\bibnamefont
			{Golubitsky}},\ }\bibfield  {title} {\enquote {\bibinfo {title}
			{Symmetry-increasing bifurcation of chaotic attractors},}\ }\href@noop {}
	{\bibfield  {journal} {\bibinfo  {journal} {Physica D}\ }\textbf {\bibinfo
			{volume} {32}},\ \bibinfo {pages} {423--436} (\bibinfo {year}
		{1988})}\BibitemShut {NoStop}%
	\bibitem [{\citenamefont {Park}, \citenamefont {Zaks},\ and\ \citenamefont
		{Kurths}(1999)}]{park1999phase}%
	\BibitemOpen
	\bibfield  {author} {\bibinfo {author} {\bibfnamefont {E.-H.}\ \bibnamefont
			{Park}}, \bibinfo {author} {\bibfnamefont {M.~A.}\ \bibnamefont {Zaks}}, \
		and\ \bibinfo {author} {\bibfnamefont {J.}~\bibnamefont {Kurths}},\
	}\bibfield  {title} {\enquote {\bibinfo {title} {Phase synchronization in the
				forced lorenz system},}\ }\href@noop {} {\bibfield  {journal} {\bibinfo
			{journal} {Phys. Rev. E}\ }\textbf {\bibinfo {volume} {60}},\ \bibinfo
		{pages} {6627} (\bibinfo {year} {1999})}\BibitemShut {NoStop}%
	\bibitem [{\citenamefont {Bhagyaraj}\ \emph {et~al.}(2023)\citenamefont
		{Bhagyaraj}, \citenamefont {Sabarathinam}, \citenamefont {Ahamed},\ and\
		\citenamefont {Thamilmaran}}]{bhagyaraj2023super}%
	\BibitemOpen
	\bibfield  {author} {\bibinfo {author} {\bibfnamefont {T.}~\bibnamefont
			{Bhagyaraj}}, \bibinfo {author} {\bibfnamefont {S.}~\bibnamefont
			{Sabarathinam}}, \bibinfo {author} {\bibfnamefont {A.~I.}\ \bibnamefont
			{Ahamed}}, \ and\ \bibinfo {author} {\bibfnamefont {K.}~\bibnamefont
			{Thamilmaran}},\ }\bibfield  {title} {\enquote {\bibinfo {title}
			{Super-extreme events in a forced bonhoeffer--van der pol oscillator},}\
	}\href@noop {} {\bibfield  {journal} {\bibinfo  {journal} {Pramana}\ }\textbf
		{\bibinfo {volume} {97}},\ \bibinfo {pages} {170} (\bibinfo {year}
		{2023})}\BibitemShut {NoStop}%
	\bibitem [{\citenamefont {Eichner}\ \emph {et~al.}(2007)\citenamefont
		{Eichner}, \citenamefont {Kantelhardt}, \citenamefont {Bunde},\ and\
		\citenamefont {Havlin}}]{eichner2007statistics}%
	\BibitemOpen
	\bibfield  {author} {\bibinfo {author} {\bibfnamefont {J.~F.}\ \bibnamefont
			{Eichner}}, \bibinfo {author} {\bibfnamefont {J.~W.}\ \bibnamefont
			{Kantelhardt}}, \bibinfo {author} {\bibfnamefont {A.}~\bibnamefont {Bunde}},
		\ and\ \bibinfo {author} {\bibfnamefont {S.}~\bibnamefont {Havlin}},\
	}\bibfield  {title} {\enquote {\bibinfo {title} {Statistics of return
				intervals in long-term correlated records},}\ }\href@noop {} {\bibfield
		{journal} {\bibinfo  {journal} {Phys. Rev. E Stat. Nonlin. Soft Matter Phys}\
		}\textbf {\bibinfo {volume} {75}},\ \bibinfo {pages} {011128} (\bibinfo
		{year} {2007})}\BibitemShut {NoStop}%
	\bibitem [{\citenamefont {Santhanam}\ and\ \citenamefont
		{Kantz}(2008)}]{santhanam2008return}%
	\BibitemOpen
	\bibfield  {author} {\bibinfo {author} {\bibfnamefont {M.}~\bibnamefont
			{Santhanam}}\ and\ \bibinfo {author} {\bibfnamefont {H.}~\bibnamefont
			{Kantz}},\ }\bibfield  {title} {\enquote {\bibinfo {title} {Return interval
				distribution of extreme events and long-term memory},}\ }\href@noop {}
	{\bibfield  {journal} {\bibinfo  {journal} {Phys. Rev. E Stat. Nonlin. Soft
				Matter Phys}\ }\textbf {\bibinfo {volume} {78}},\ \bibinfo {pages} {051113}
		(\bibinfo {year} {2008})}\BibitemShut {NoStop}%
	\bibitem [{\citenamefont {Blender}, \citenamefont {Fraedrich},\ and\
		\citenamefont {Sienz}(2008)}]{blender2008extreme}%
	\BibitemOpen
	\bibfield  {author} {\bibinfo {author} {\bibfnamefont {R.}~\bibnamefont
			{Blender}}, \bibinfo {author} {\bibfnamefont {K.}~\bibnamefont {Fraedrich}},
		\ and\ \bibinfo {author} {\bibfnamefont {F.}~\bibnamefont {Sienz}},\
	}\bibfield  {title} {\enquote {\bibinfo {title} {Extreme event return times
				in long-term memory processes near 1/f},}\ }\href@noop {} {\bibfield
		{journal} {\bibinfo  {journal} {Nonlinear Process. Geophys.}\ }\textbf
		{\bibinfo {volume} {15}},\ \bibinfo {pages} {557--565} (\bibinfo {year}
		{2008})}\BibitemShut {NoStop}%
	\bibitem [{\citenamefont {Santhanam}\ and\ \citenamefont
		{Kantz}(2005)}]{santhanam2005long}%
	\BibitemOpen
	\bibfield  {author} {\bibinfo {author} {\bibfnamefont {M.}~\bibnamefont
			{Santhanam}}\ and\ \bibinfo {author} {\bibfnamefont {H.}~\bibnamefont
			{Kantz}},\ }\bibfield  {title} {\enquote {\bibinfo {title} {Long-range
				correlations and rare events in boundary layer wind fields},}\ }\href@noop {}
	{\bibfield  {journal} {\bibinfo  {journal} {Phys. A: Stat. Mech. Appl.}\
		}\textbf {\bibinfo {volume} {345}},\ \bibinfo {pages} {713--721} (\bibinfo
		{year} {2005})}\BibitemShut {NoStop}%
	\bibitem [{\citenamefont {Hochreiter}\ and\ \citenamefont
		{Schmidhuber}(1997)}]{lstm}%
	\BibitemOpen
	\bibfield  {author} {\bibinfo {author} {\bibfnamefont {S.}~\bibnamefont
			{Hochreiter}}\ and\ \bibinfo {author} {\bibfnamefont {J.}~\bibnamefont
			{Schmidhuber}},\ }\bibfield  {title} {\enquote {\bibinfo {title} {{Long
					Short-Term Memory}},}\ }\href@noop {} {\bibfield  {journal} {\bibinfo
			{journal} {Neural Comput}\ }\textbf {\bibinfo {volume} {9}},\ \bibinfo
		{pages} {1735--1780} (\bibinfo {year} {1997})}\BibitemShut {NoStop}%
\end{thebibliography}

%

\end{document}